\documentclass[10.5pt,preprint]{aastex}

\shorttitle{Black Holes in Dwarf Galaxies}
\shortauthors{Moran et al.}

\begin{document}

\title{Black Holes at the Centers of Nearby Dwarf Galaxies}

\author{Edward C.\ Moran\altaffilmark{1}, Karlen Shahinyan\altaffilmark{1,2},
        Hannah R.\ Sugarman\altaffilmark{1}, Darik O.\ V\'elez\altaffilmark{1},
        \\and Michael Eracleous\altaffilmark{3,4,5}}

\altaffiltext{1}{Astronomy Department, Wesleyan University, Middletown, CT
                 06459.}
\altaffiltext{2}{Institute for Astrophysics, University of
                 Minnesota, 116 Church St.\ S.E., Minneapolis, MN 55455.}
\altaffiltext{3}{Department of Astronomy and Astrophysics, and Institute for
                 Gravitation and the Cosmos, The Pennsylvania State University,
                 525 Davey Lab, University Park, PA 16802.}
\altaffiltext{4}{Center for Relativistic Astrophysics, Georgia Institute of
                 Technology, Atlanta, GA 30332.}
\altaffiltext{5}{Department of Astronomy, University of Washington, Seattle,
                 WA 98195.}

\begin{abstract}
Using a distance-limited portion of the Sloan Digital Sky Survey (SDSS) Data
Release 7, we have identified 28 active galactic nuclei (AGNs) in nearby
($d \le 80$ Mpc) low-mass, low-luminosity dwarf galaxies.  The accreting
objects at the galaxy centers are expected to be intermediate-mass black
holes (IMBHs) with $M_{\rm BH} \le 10^6$~$M_{\odot}$.  The AGNs were selected
using several optical emission-line diagnostics after careful modeling of
the continuum present in the spectra.  We have limited our survey to objects
with spectral characteristics similar to those of Seyfert nuclei, excluding
emission-line galaxies with ambiguous spectra that could be powered by
stellar processes.  The host galaxies in our sample are thus the least
massive objects in the very local universe certain to contain central black
holes.  Our sample is dominated by narrow-line (type 2) AGNs, and it appears
to have a much lower fraction of broad-line objects than that observed
for luminous, optically selected Seyfert galaxies.  Given our focus
on the nearest objects included in the SDSS, our survey is more sensitive
to low-luminosity emission than previous optical searches
for AGNs in low-mass galaxies.  The [\ion{O}{3}] $\lambda 5007$ luminosities
of the Seyfert nuclei in our sample have a median value of $L_{5007} =
2 \times 10^5$ $L_{\odot}$ and extend down to $\sim 10^4$ $L_{\odot}$.  Using
published data for broad-line IMBH candidates, we have derived an
[\ion{O}{3}] bolometric correction of $\log (L_{\rm bol}/L_{5007}) = 3.0 \pm
0.3$, which is significantly lower than values obtained for
high-luminosity AGNs.  Applying this correction to our sample, we obtain
minimum black-hole mass estimates that fall mainly in the $10^3$ $M_{\odot}$
-- $10^4$ $M{_\odot}$ range, which is roughly where the predicted mass
functions for different black-hole seed formation scenarios overlap the
most.  In the stellar mass range that includes the bulk of the AGN host
galaxies in our sample, we derive a lower limit on the AGN fraction of a
few percent, indicating that active nuclei in dwarf galaxies are not as
rare as previously thought.
\end{abstract}


\section{Introduction}

Numerous studies over the past two decades have revealed correlations
between the masses of supermassive black holes ($M_{\rm BH}$) and the
large-scale properties of the galaxies within which they reside,
such as the luminosity (e.g., Kormendy \& Richstone 1995; Bentz et al.\
2009), mass (Marconi \& Hunt 2003), and velocity dispersion ($\sigma_{\star}$;
Gebhardt et al.\ 2000; Ferrarese \& Merritt 2000; Tremaine et al.\ 2002;
G\"ultekin et al.\ 2009) of the stellar bulge.  Based on these correlations,
it is now widely presumed that supermassive black holes and the bulges of
luminous galaxies grow and evolve in a coordinated manner via a single,
common process.  The case for the leading candidate, galaxy merging, is
supported by simulations, which demonstrate that nuclear black hole growth
occurs primarily through merger-induced accretion (Volonteri et al.\ 2003;
Di Matteo et al.\ 2005; Hopkins et al.\ 2005).  However, the matter is far
from settled.  In addition to the fact that some nearby late-type galaxies
may lack central black holes (Gebhardt et al.\ 2001; Valluri et al.\ 2005),
a number of critical issues remain unresolved.  For example, do the scaling
relations observed for bulge-dominated galaxies and supermassive black holes
(e.g., $M_{\rm BH}$--$\sigma_{\star}$) apply in the low-mass regime?  What are
the origins of the black hole ``seeds'' that formed at earlier times, and
what were their initial masses?  Without answers to these questions, we lack
a full understanding of the coordinated growth of black holes and galaxies,
and, thus, a fundamental aspect of galaxy evolution.

Studies of nearby low-mass galaxies can provide key insight into these
issues.  Low-mass galaxies are likely to have had quiet merger histories,
and the black holes at their centers, if present, are expected to be in
the intermediate-mass range (i.e., $M_{\rm BH} = 10^3$--$10^6\, M_{\odot}$).
Thus, the discovery of dwarf galaxies with massive central black holes
affords the opportunity to explore black-hole/galaxy scaling relations in
objects similar to the precursors of today's large galaxies.  Furthermore,
as discussed by Volonteri et al.\ (2008), the present-epoch ``occupation
fraction'' of black holes in galaxies --- particularly those at the lowset
masses --- can be used to discriminate between seed formation scenarios,
i.e., ``light'' seeds from Population III stellar remnants (Bromm et al.\
1999) and ``heavy'' seeds formed from the collapse of metal-free gas in
primordial galaxies (Lodato \& Natarajan 2006).  But in order for such
investigations to succeed, two things are essential:\ (1) a determination
of the demographics of massive black holes in the local universe, and (2)
an accurate census of the properties of nearby galaxies that contain central
black holes.

Much progress has been made recently in the identification and
characterization of intermediate-mass black hole (IMBH) candidates and
their host galaxies (Greene \& Ho 2004, 2007b; Barth et al.\ 2004, 2005,
2008; Dong et al.\ 2007, 2012; Xiao et al.\ 2011; Jiang et al.\ 2011).
However, as discussed by Greene \& Ho (2007a), the search techniques
used to select the objects for these studies have yielded samples that
suffer to some degree from luminosity bias, resulting in incompleteness
for both black holes and their hosts at low masses.  In this paper, we
report on the first of a series of projects designed to improve our
understanding of the relationship between black holes and galaxies in
the low-mass regime.  Our approach and data selection methods are
outlined in \S~2.  Host-galaxy photometry of our sample is described in
\S~3, and in \S~4 we discuss the processing and analysis of the available
nuclear spectroscopic data.  Our survey has revealed a number of dwarf
galaxies that contain a central black hole (Moran 2010); their properties
are presented in \S~5.  The paper concludes with a summary of our findings.

\section{Sample Definition}

\subsection{A Distance-Limited Approach}

Due to the small physical size of the sphere of influence of an IMBH and
the limited resolving power of available instrumentation, dynamical methods
for discovering (or setting upper limits on the masses of) such objects
(Gebhardt et al.\ 2001; Valluri et al.\ 2005; Lora et al.\ 2009; Seth et
al.\ 2010; Jardel \& Gebhardt 2012; Neumayer \& Walcher 2012) can only be
employed for the very nearest galaxies (Ferrarese 2004).  Thus, searches
for IMBHs --- even in the local universe --- must rely on the detection
of nuclear activity in a galaxy for evidence that a black hole is present.
Unfortunately, owing to the fact that accretion luminosity depends in part
on $M_{\rm BH}$, active galactic nuclei (AGNs) with low-mass black holes
tend to be faint.  As a result, current samples of IMBH candidates are
relatively small.  The original and best-studied case is NGC 4395, a dwarf
spiral galaxy that possesses a broad emission-line (type 1) nucleus
(Filippenko \& Sargent 1989) and a $\sim 10^5$ $M_{\odot}$ black hole
(Filippenko \& Ho 2003; Peterson et al.\ 2005).  The other well-studied
example is POX 52 (Barth et al.\ 2004; Thornton et al.\ 2008), which,
despite being similar to NGC 4395 in terms of its black-hole mass and the
optical and X-ray properties of its nucleus, is located in a dwarf
spheroidal galaxy.

In recent years, a few hundred additional IMBH candidates have been
identified using spectroscopic data from the Sloan Digital Sky Survey (SDSS;
see Greene \& Ho 2004, 2007b; Dong et al.\ 2007, 2012; Barth et al.\ 2008;
Reines et al.\ 2013).  The samples of Greene \& Ho (2004; 2007b) and Dong
et al.\ (2012) were limited to broad-line AGNs, since black-hole mass
estimates can be made for them.  The Barth et al.\ (2008) and Reines et
al.\ (2013) surveys targeted lower mass host galaxies, yielding samples
dominated by narrow-line (type 2) AGNs.  Although all of these studies
were successful at demonstrating the existence of IMBHs in other galaxies,
the relatively high signal-to-noise ({\sl S/N}) ratio thresholds they
required, coupled with the fact that the SDSS is essentially a flux-limited
survey, introduced luminosity bias into their AGN samples.  For example,
nuclear luminosity (indicated by the strength of the [\ion{O}{3}] $\lambda
5007$ emission line) appears to correlate with distance in the Greene \& Ho
(2004) and Barth et al.\ (2008) samples.  This arises in circumstances where
low-luminosity objects are too faint to be detected at large distances,
and high-luminosity objects are too rare to be found within a small local
volume.  The net effect for the IMBH samples is that their nuclear
luminosity distributions, compared to the [\ion{O}{3}] luminosity function,
are shifted to higher values.

Evidently, the host galaxies of emission-line selected IMBH candidates in
the first SDSS surveys are biased to higher luminosities as well.  As pointed
out by Greene \& Ho (2007a) and Barth et al.\ (2008), galaxies must be
sufficently bright to be targeted for spectroscopy in the SDSS.  Given the
distances of the objects in these samples (the median redshift of the
Greene \& Ho sample is $z \approx 0.10$), this translates to relatively
high luminosities (and masses) for the IMBH host galaxies.  The objects
included in the Greene \& Ho (2004; 2007b) and Barth et al.\ (2008) surveys
are typically $\sim$1 mag fainter than $L^{*}$, but very few are faint enough
to qualify them as dwarf galaxies similar to NGC~4395 and POX 52.

The presence of luminosity bias in existing IMBH samples could mean that the
low-mass limits for central black holes and/or their host galaxies have not
yet been reached.  If this is the case, we lack crucial information about the
very objects that could provide the greatest leverage (via their black hole
occupation fraction) on the massive black-hole seed formation mechanism.
The goal of the project described here, therefore, is to investigate whether
or not weaker AGNs can be identified in even less-luminous galaxies. The
success of the surveys cited above demonstrates that even at low values of
$M_{\rm BH}$ emission-line diagnostics remain a powerful tool for identifying
AGNs.  We have therefore adopted this strategy as well.
However, the approach of our survey differs in an important way:\  rather
than selecting objects based on their spectral properties, we instead
assemble a sample of all potential AGN host galaxies within a certain
distance limit and {\it then\/} search each for emission-line evidence of
black-hole accretion.  Our focus is on the nearest objects (1) to enable
the detection of weak active nuclei in the faintest (and least massive) host
galaxies, and (2) to maximize the sensitivity of any follow-up observations.
As we discuss below, our sample provides the best coverage to date of the
lower portions of both the nuclear and host-galaxy luminosity functions.

\subsection{Data Selection}

As with the prior emission-line searches for IMBH candidates listed above,
the data for our survey are drawn from the SDSS.  In its seventh data release
(DR7; Abazajian et al.\ 2009), the spectroscopic portion of the SDSS covers
$\sim 8200$ deg$^2$ of sky, primarily in the North Galactic Cap.  We began
by selecting all objects in the DR7 that have (1) extragalactic spectral
classifications (i.e., star and sky spectra were omitted, but objects with
``unknown'' classifications were included) and (2) heliocentric recessional
velocities of $v_{\rm r} \le 5300$ km~s$^{-1}$ (i.e., $z \le 0.0177$).  The
velocity limit was adopted based on our rough estimates that objects like
NGC~4395 would just be included in the SDSS at the corresponding distance
(galaxies must be brighter than $r = 17.77$ mag for SDSS spectroscopy) or
that we would be able to observe them with medium-aperture telescopes if
they were not.  We corrected the recessional velocities for infall due to
the Virgo cluster and Great Attractor following Mould et al.\ (2000), with
an adjustment for the Local Group component described by Karachentsev \&
Makarov (1996).  Physical distances to the objects were derived from their
corrected velocities by applying the Hubble Law with
$H_0 = 73$ km~s$^{-1}$~Mpc$^{-1}$.
For the majority of the sample, the adopted velocity limit corresponds to
a maximum distance of $\sim 80$ Mpc.  The distances of objects with the
lowest redshifts and those that lie in the direction of the Virgo cluster
have the greatest uncertainty; as discussed below in \S~5.1, the distances
of all AGNs in our sample have been closely scrutinized to ensure that our
assessments of their luminosities and masses are as accurate as possible.

This selection produced an initial sample of over $10^4$ spectra.  However,
a problem encountered with nearby galaxies is that, quite frequently,
extranuclear regions have been observed spectroscopically by the SDSS rather
than (or in addition to) the nucleus or central region of the galaxy.  We
decided that visual inspection of SDSS images indicating the placement of
the spectral fiber was the most reliable (though certainly not the most
expediant) method of identifying and eliminating extranuclear spectra.  The
process was straightforward for the most part, but for irregular galaxies
and small, knotty galaxies at large distances it was sometimes unclear if
an object's ``central'' region had been observed.  These were evaluated
case-by-case; in the end, we kept the spectra only when we felt that an
attempt to reobserve the galaxy would not result in a better characterization
of its nucleus.

In total, the filtering process yielded a final sample of 9,526 galaxies.
We then acquired their photometric data from the SDSS to identify low-mass
galaxies (\S~3), and processed and analyzed their spectra to search for
nuclear activity in the objects (\S~4).

\section{SDSS Photometry}

Accurate photometry is critical for this project in order to establish the
properties of any black-hole host galaxies we uncover and to characterize
the parent sample as a whole.  For the vast majority of objects in our sample,
the photometric data available from the SDSS provide a reliable measure of a
galaxy's brightness.  Whenever possible, we have employed the SDSS Petrosian
magnitudes, which are derived from aperture photometry; they are independent
of surface brightness modeling and afford the most direct comparison between
our sample and the samples examined in previous IMBH surveys (e.g., Greene
\& Ho 2004; Barth et al.\ 2008).

We note, however, that in some instances errors are present in the SDSS
photometry.  For our sample of nearby galaxies, this seems to involve
primarily two types of objects:\ very extended galaxies, and less extended
objects that lie close in projection to a bright foreground star. Photometry
problems can be revealed by examining the SDSS photometric pipeline's object
flags and/or by noting significant discrepancies when various types of SDSS
magnitudes are compared, e.g., the Petrosian and ``model'' magnitudes
(which are obtained by fitting an exponential or de~Vaucouleurs model to
a galaxy's surface brightness profile), or the DR7 magnitudes and those
recalculated as part of the DR8 (Aihara et al.\ 2011).  Unfortunately, it is
not automatically clear which set of measurements (if any) should be preferred
when a problem is suspected.  Therefore, as discussed in \S~5.1, we carefully
scrutinized the photometry data available for any objects identified as AGNs.

An accurate characterization of our overall galaxy sample is possible if we
temporarily ignore objects with suspected photometry problems and those with
the greatest relative uncertainties in their distances (i.e., objects with
the lowest redshifts [$z < 0.003$] and potential members of the Virgo cluster;
see Mould et al.\ 2000).  Such exclusions reduce our sample by only $\sim
12$\%.  The absolute magnitude and stellar mass distributions for the 7415
remaining galaxies are displayed in Figure~1.  Absolute magnitudes were
computed using our distance estimates (\S~2.2) and the SDSS $g$-band magnitudes
(which are best for comparison to the Greene \& Ho and Barth et al.\ samples).
We estimated the stellar masses of the galaxies following the method of Bell
et al.\ (2003), which combines a galaxy's luminosity with a mass-to-light
ratio derived from a color measurement.  Our calculations employ the $g-r$
colors and $i$-band luminosities.  As Figure~1 indicates, our distance-limited
sample is dominated by dwarf galaxies fainter than $M_g = -18$ and less massive
than $10^{10}$~$M_{\odot}$ --- the very population we wish to search for
evidence of nuclear activity.

\vskip -1.4truein
\begin{figure}[!htb]
\hskip -0.3truein
\includegraphics[width=0.60\textwidth]{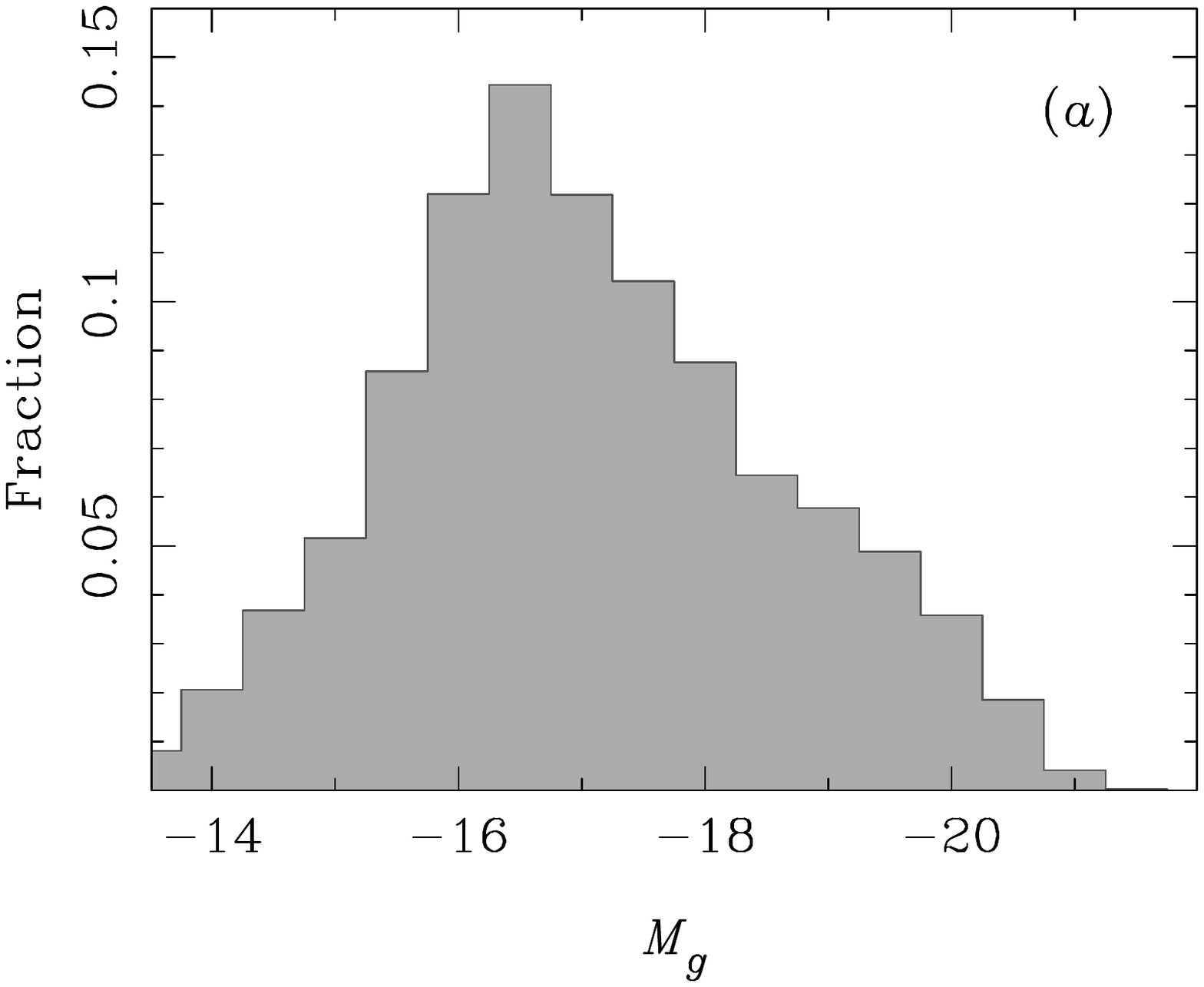}
\vskip -5.06truein
\hskip 2.95truein
\includegraphics[width=0.60\textwidth]{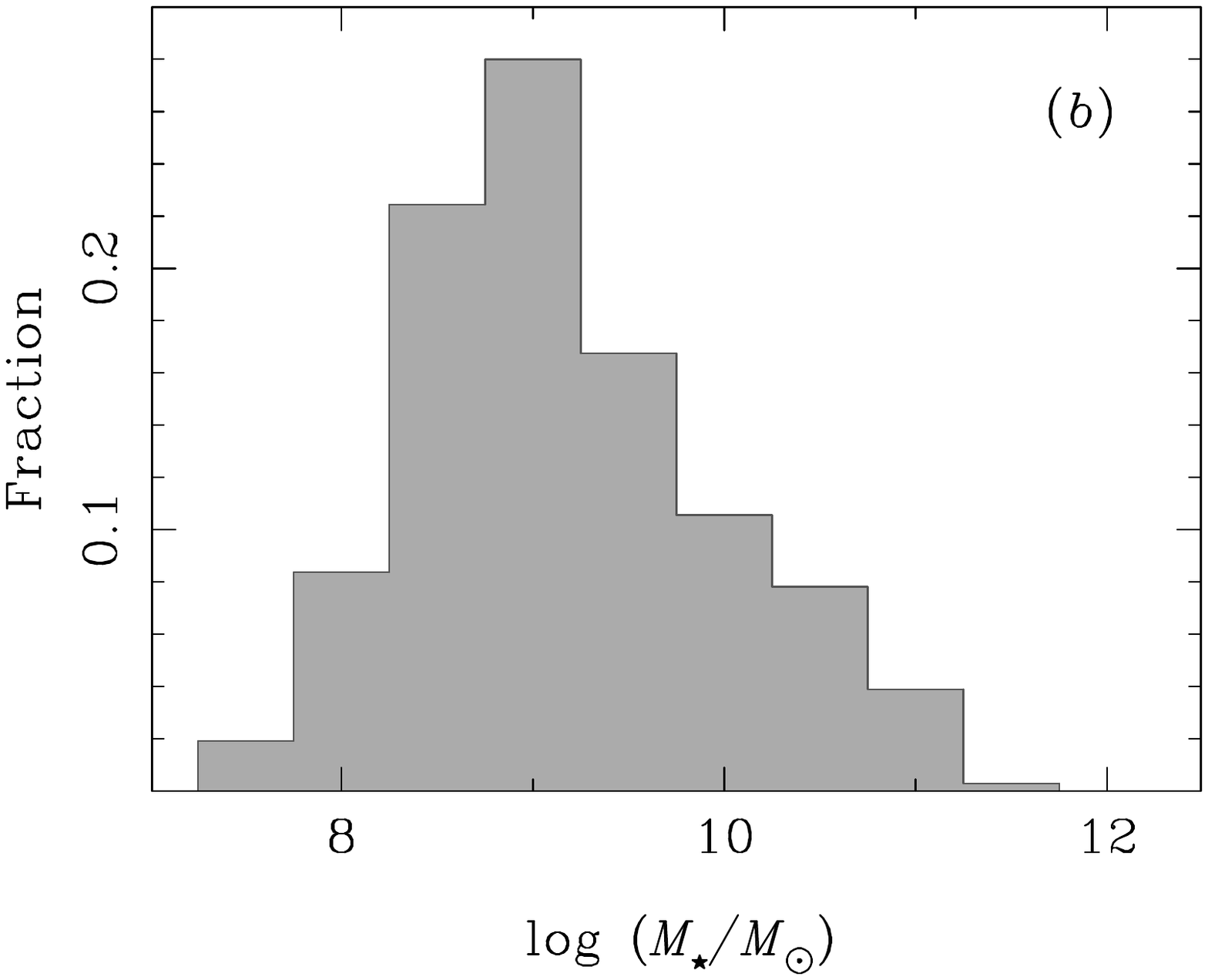}
\end{figure}

\vskip -1.1truein
{\narrower\noindent\small Fig.~1.---($a$) Absolute $g$ magnitude and ($b$)
stellar mass distributions for our SDSS DR7 sample of nearby galaxies.\par}

\section{SDSS Spectroscopy}

Optical spectra of the galaxies in our sample were acquired from the SDSS
DR7 database (Abazajian et al.\ 2009).  The spectra, obtained using fibers
that provide an effective $3''$ diameter aperture, span a wavelength range
of $\sim$ 3800--9200 \AA\ at a resolving power of
$\lambda / \Delta\lambda \approx 2000$.  Two types of analysis were performed
in order to evaluate the nuclear activity of each object:\ (1) subtraction
of the stellar continuum present in the raw spectrum, and (2) measurement
of the fluxes of the important diagnostic emission lines.

\subsection{Continuum Modeling}

Regardless of whether the galaxies in our sample are AGNs, their nuclear
spectra are almost always dominated by starlight from their central regions.
The superposition of absorption features in the stellar continuum and
nebular emission lines principally affects the apparent strengths of the
narrow Balmer lines.  However, depending on the stellar population present
and the relative strength of the line emission, the apparent fluxes of some
important diagnostic forbidden lines can be affected as well.  In addition,
when host-galaxy emission dominates, it can mask the presence of weak, broad
emission lines associated with an active nucleus.  Thus, as discussed by
Ho (2008), careful subtraction of the stellar continuum is essential for
an accurate emission-line assessment of an object's nuclear activity.

Our initial analysis employed the continuum-fitting code developed by
Eracleous \& Halpern (2001).  After masking out the locations of strong
emission lines, the program constructs a continuum model for each galaxy
spectrum by fitting a linear combination of starlight templates and a
non-stellar power-law component.  The emission-line flux ratios in
the residual spectrum (obtained by subtracting the continuum model from the
data) are then used to classify the galaxies (Veilleux \& Osterbrock 1987;
Kewley et al.\ 2006).  The program was designed to use absorption-line
galaxy spectra as templates, as these often accurately represent the
stellar populations and velocity dispersions present in the centers of
emission-line galaxies.  A search for suitable templates among the SDSS
DR7 spectra obtained for this project yielded a set of over 30 spectra
that have very high {\sl S/N\/} and appear to be devoid of emission lines.
Using these templates, we were able to identify the majority of AGNs
located in the higher mass galaxies from our sample.  However, two concerns
arose regarding the effectiveness of this approach for weak-lined dwarf
galaxies.  First, many dwarf galaxies have bluish stellar continua with
effective spectral types of G, F, or even A stars.  High-quality spectra
of nearby galaxies that have similar colors but no emission lines are
extremely rare in the DR7, so our template set was very incomplete for
such objects.  Second, many of the best templates we selected were associated
with fairly massive galaxies whose stellar velocity dispersions are larger
than those of dwarf galaxies, which might lead to improper subtractions in
the vicinities of important diagnostic emission lines.

To address these issues, we also analyzed the data with the GANDALF package
(Sarzi et al.\ 2006; Sarzi et al.\ 2007).  GANDALF operates in two stages.
Similar to the approach described above, it first obtains an initial model
for the continuum by masking off emission lines and then using the pPXF
software (Cappellari \& Emsellem 2004) to fit a linear combination of
starlight templates that has been multiplied by a low-order polynomial (to
account for reddening and flux-calibration differences between the object
and template spectra)  {\it and\/} broadened to match the line-of-sight
velocity dispersion of the galaxy.  GANDALF then lifts the masks and fits
the emission lines simultaneously with the continuum.  We used synthetic
stellar population (SSP) models based on v.~9.1 of the MILES stellar library
as starlight templates (Vazdekis et al.\ 2010); for a given prescription of
the initial mass function (IMF) and metallicity, these models are calculated
for 50 ages between 0.063 Gyr and 17.8 Gyr.  The spectral resolution of the
templates we used is $\sigma = 64$ km~s$^{-1}$.  
GANDALF fits are generally excellent, with values of $\chi^2$ per degree
of freedom of $\sim$ unity or less.  Tests using a set of $\sim 30$ galaxy
spectra with a range of emission-line strengths indicated that the two
continuum-subtraction programs yield similar results for objects with
strong emission lines, but for weak-lined objects we generally obtained
smoother residuals and hence more reliable emission-line fluxes with GANDALF.
An example of continuum fitting with GANDALF is displayed in Figure~2.
For analysis of the full sample, we employed MILES templates generated for a
Kroupa (2001) universal IMF and solar metallicity; these yielded the best
results for a test set of low-mass ($\sim 10^9$ $M_{\odot}$) galaxies.

\begin{figure}[!htb]
\vskip -0.08truein
\begin{center}
\includegraphics[width=0.87\textwidth]{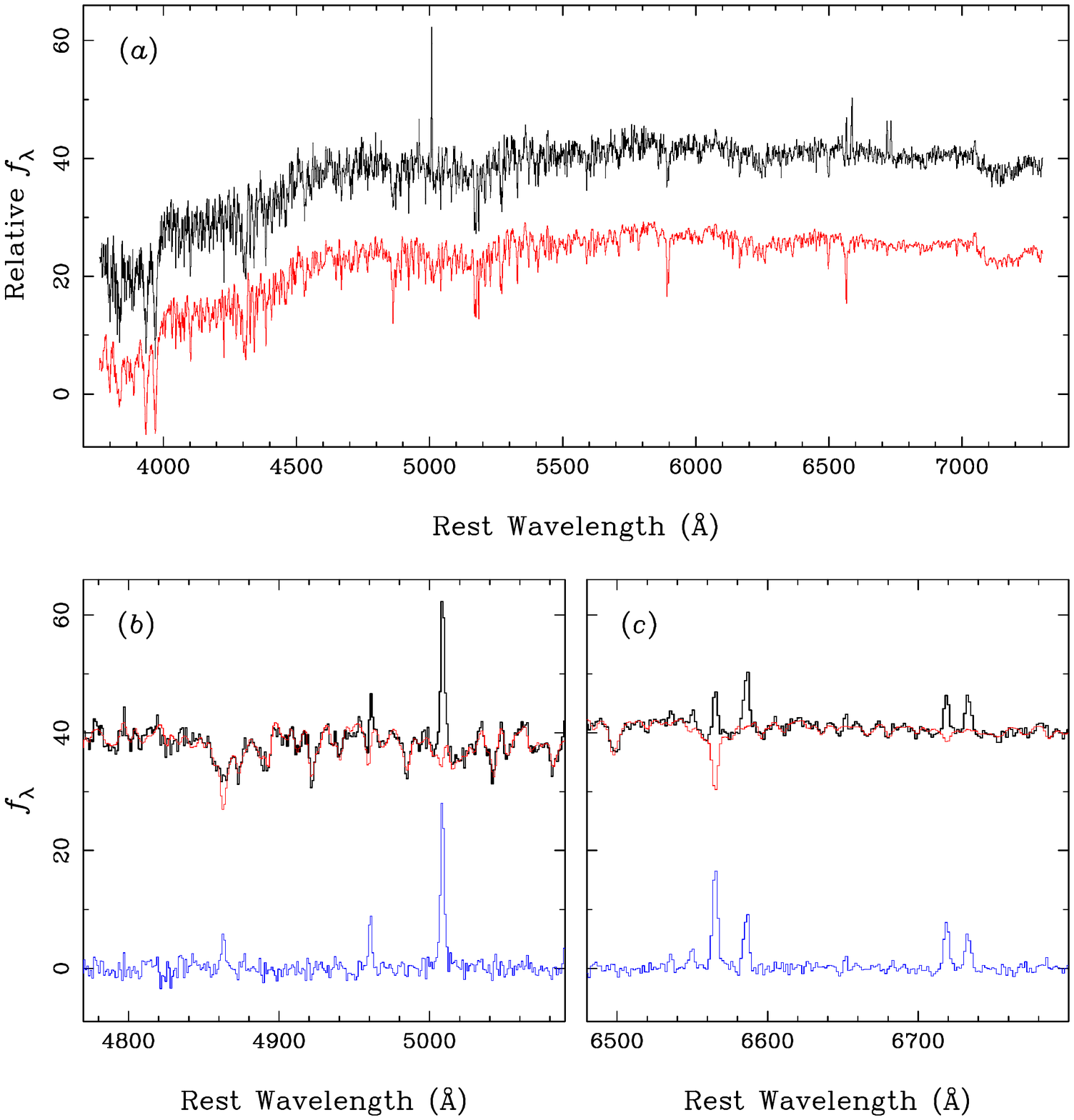}
\end{center}
\end{figure}
\vskip -1.8truein
{\narrower\noindent\small Fig.~2.---Continuum modeling of the spectrum
of J0802+1032, which falls below the
median for our dwarf galaxy AGN sample in terms its overall $S/N$ ratio
and emission-line equivalent widths.  In ($a$), the original SDSS spectrum
(black) and best-fit continuum from GANDALF (red, offset for clarity) are
displayed.  In the lower panels, the continuum fit (red) is overlaid on the
original spectrum (black) in the vicinity of ($b$) H$\beta$ $\lambda 4861$
(including [\ion{O}{3}] $\lambda\lambda 4959,5007$) and ($c$) H$\alpha$ 
$\lambda 6563$ (including [\ion{N}{2}] $\lambda\lambda 6548,6583$ and
[\ion{S}{2}] $\lambda\lambda 6716,6731$).  The emission-line flux ratios
revealed in the residuals from the continuum subtraction (blue) indicate
the presence of an active nucleus.\par}

\clearpage
\subsection{Emission-Line Measurements}

As mentioned above, GANDALF fits the emission lines simultaneously with
the continuum, so line fluxes (and fully propagated errors associated with
the fits) are obtained automatically.
This is achieved with a set of user-specified emission-line constraints
that allow one to define the profiles and relative rest-frame wavelengths
of lines, fix the flux ratios of multiplets (e.g., [\ion{O}{3}]
$\lambda\lambda 4959,5007$), and associate the velocity widths of various
lines.  For the analysis of the full sample, we (a) assumed that the lines
have Gaussian profiles and (b) tied the velocity widths of forbidden lines
to a common value that is allowed to differ from that of the permitted
lines (H and He).  This set-up works well for the majority of narrow
emission-line galaxies.

However, as discussed by Oh et al.\ (2011), GANDALF does not always achieve
acceptable fits for SDSS spectra.  Moreover, no single set of assumptions
about the continuum templates or emission-line properties is appropriate
for all objects.  Inaccuracies in automatically measured line fluxes can
occur whenever the continuum or line profiles are not properly modeled,
especially for weak-lined objects.  Thus, we have taken additional steps to
verify the accuracy of the line-flux measurements.  Our main goal is to
ensure that no potential AGN candidates have been overlooked.

To do this, we visually inspected the continuum fitting results for the
entire sample.  First, we constructed a residual spectrum for each object by
subtracting the continuum model obtained with GANDALF from the original data.
Ideally, this should contain any emission lines present and a continuum
with a mean of zero that reflects the noise in the raw data and
errors associated with the fit.  We examined each residual spectrum using
three-paneled plots that show the full wavelength range and close-ups of
the H$\alpha$ and H$\beta$ regions.  Poor continuum or emission-line fits
are easily spotted this way, and the experienced eye can quicky recognize
subtle AGN features that would be difficult to identify
as completely and efficiently in an automated manner.  We also examined a
``noise'' spectrum for each galaxy constructed by subtracting the total
model (continuum plus emission lines) from the data.  Residuals in the
noise spectrum at the positions of emission lines indicate objects with
non-Gaussian line profiles.

We adopted GANDALF line-flux measurements for objects whose residual spectrum
is reasonably flat in the continuum and whose noise spectrum exhibits no
strong features at the positions of the diagnostic emission lines.  If these
conditions were not met, or if something in the residual spectrum suggested
that an AGN may be present, we analyzed the data further.  For objects with
poor continuum fits, we refitted the spectra in GANDALF using templates with
non-solar metallicity, including a non-stellar power law ($F \propto
\nu^{-\alpha}$, with $\alpha = 1 - 2$) as one of the templates, and/or
including broad Balmer-line components in the emission-line fits.  In some
instances, increasing the order of the multiplicative polynomial applied
to the templates was necessary to eliminate low-frequency wiggles in the
residual continuum.  For objects with non-Gaussian line profiles, we either
added more emission components in follow-up GANDALF fits or used the
{\it splot\/} task in IRAF to remeasure the fluxes of lines by directly
integrating residual spectrum.  The latter approach was best for narrow-line
objects with Lorentzian or Voigt line profiles.  And finally, for AGN
candidates selected visually, we carefully reexamined all of the information
contained in the spectrum (e.g., the velocity widths of various lines or the
presence of lines with high ionization potentials) and remeasured the
emission-line fluxes.

\subsection{Upper and Lower Limits}

The visual inspection of the continuum-fitting results described above
revealed some potential AGN candidates that lack detections of one or more
of the lines useful for diagnosing nuclear activity.  For example, we noticed
a number of objects with strong [\ion{O}{3}] $\lambda 5007$ emission for
which the weaker H$\beta$ line is not detected.  For these objects, we
measured the flux in 40--50 independent bins on either side of [\ion{O}{3}]
+ H$\beta$ in the residual spectrum.  The widths of the bins were set to
be roughly equal to the width of [\ion{O}{3}] $\lambda 5007$ near its base.
We then computed H$\beta$ upper limits as three times the RMS of the flux
in the continuum bins.  As a safety check, we overplotted the results on
the residual spectrum to ensure that the derived limits are consistent with
the noise levels in the spectra.  Two examples are shown in Figure~3.

\vskip -0.95truein
\begin{figure}[!htb]
\vskip 0.05truein
\hskip -0.15truein
\includegraphics[trim=0 850 0 850,clip,width=\textwidth]{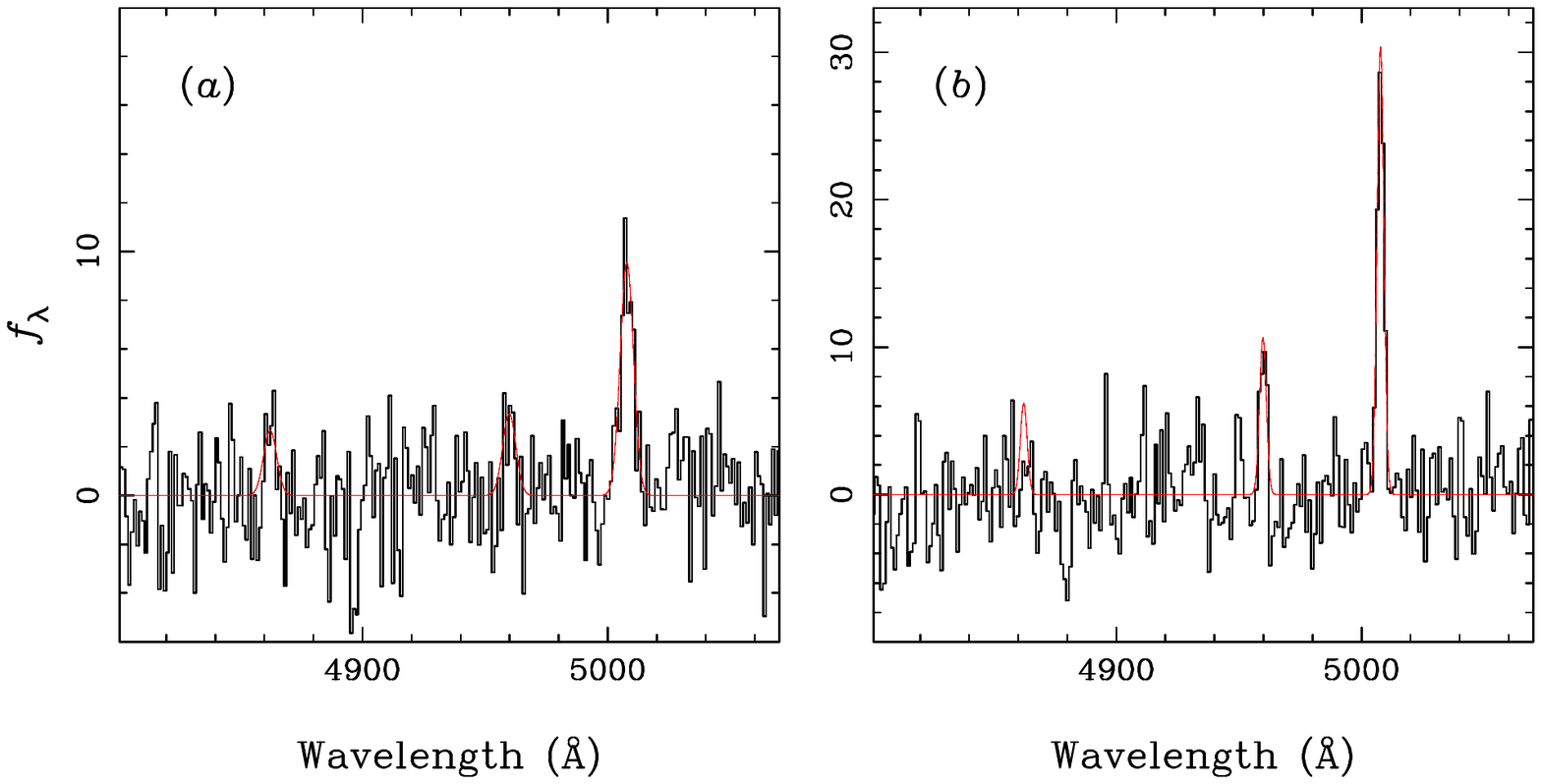}
\end{figure}

\vskip 4.2truein
{\narrower\noindent\small Fig.~3.---Determination of H$\beta$ upper limits
for two weak-lined objects, ($a$) J0932+3141 and ($b$) J1538+1204.  Overlaid
on each continuum-subtracted spectrum is an emission-line model consisting
of the best-fit Gaussian to the [\ion{O}{3}] $\lambda 5007$ line, the
[\ion{O}{3}] $\lambda 4959$ line expected from the $\lambda 5007$ fit,
and the measured 3~$\sigma$ upper limit of H$\beta$, which is assumed to
have the same velocity width as [\ion{O}{3}].  In both cases, the model
accurately represents the flux observed at $\lambda 4959$.  Some H$\beta$
flux appears to be present in the spectrum of J0932+3141, though the line
is not formally detected.\par}

\clearpage
In addition, we estimated lower limits on the combined
[\ion{S}{2}] $\lambda 6716 + \lambda 6731$ flux for potential AGN candidates
in which the former line was detected but the latter line was not.  A lower
limit on their sum was obtained by assuming $F(\lambda 6716)/F(\lambda 6731)
= 1.4$, which is the maximum ratio expected in low-density photoionized
nebulae (Osterbrock 1989).

\section{Results}

\subsection{Identifying AGNs in Low-Mass Galaxies}

We used the following approach to construct our sample of AGNs in low-mass
galaxies: (1) we culled all objects that, based on the available emission-line
evidence, appear to be AGNs, (2) we verified the accuracy of the distances and
photometry for all AGN candidates, and (3) we imposed a maximum value for the
stellar mass of the AGN hosts.

Classification of the nuclear activity of the galaxies was made primarily
on the basis of their locations on standard flux-ratio diagnostic diagrams,
i.e., [\ion{O}{3}] $\lambda 5007$/H$\beta$ vs.\
       [\ion{N}{2}] $\lambda 6583$/H$\alpha$,
       [\ion{O}{3}] $\lambda 5007$/H$\beta$ vs.\
       [\ion{S}{2}] $\lambda\lambda 6716,6731$/H$\alpha$, and
       [\ion{O}{3}] $\lambda 5007$/H$\beta$ vs.\
       [\ion{O}{1}] $\lambda 6300$/H$\alpha$
(Veilleux \& Osterbrock 1987).  In most cases, we required that AGNs (a) lie
across the empirical ``maximum starburst line'' (Kewley et al.\ 2006) from
the locus of \ion{H}{2} galaxies on each plot, and (b) have Seyfert levels
of ionization with [\ion{O}{3}] $\lambda$5007/H$\beta$ ratios in excess of
3.  Although LINERs and similar objects with [\ion{O}{3}]/H$\beta < 3$
can arise via photoionization by a dilute non-stellar continuum from an
AGN (Ferland \& Netzer 1983; Halpern \& Steiner 1983; Filippenko \& Halpern
1984; Ho et al.\ 2003), their observed line ratios can also result from
purely stellar processes (Shields 1992; Filippenko \& Terlevich 1992;
Binette et al.\ 1994; Dopita \& Sutherland 1995).  Additionally, recent
work has shown that even if an AGN is present, black-hole accretion is
insufficient to power the emission lines associated with many LINERs
(Eracleous et al.\ 2010; Yan \& Blanton 2012).  Thus, a LINER spectrum does
not automatically point to the presence of a massive black hole, and we have
excluded all LINER-like objects with low $\lambda$5007/H$\beta$ ratios unless
their spectra contain additional evidence in support of an AGN interpretation
(e.g., high-ionization emission lines such as \ion{He}{2} $\lambda 4686$,
[\ion{Fe}{7}] $\lambda 6087$, and [\ion{Fe}{10}] $\lambda 6375$, which are
indicative of a hard ionizing continuum).
In general, the high [\ion{O}{3}]/H$\beta$ condition set here
ensures that the galaxies we select for further study are powered (at least
in part) by black-hole accretion and that our sample, as a whole, is
unconfused by non-AGNs --- a prudent choice, given that the relationship
between galaxies and black holes at low masses is not well understood.

For each AGN meeting the above criteria, we reexamined the estimated distance
and SDSS photometry to select objects with stellar masses consistent with
those of dwarf galaxies.  Concerning distances, we searched the NASA/IPAC
Extragalactic Database (NED) for objects with redshift-independent assessments
of their distance (e.g., via the Tully-Fisher method), and we investigated
whether objects presumed to be members of the Virgo cluster ($d \approx 14$
Mpc) might instead be background galaxies.  As noted above in \S~3, we used
the DR7 Petrosian magnitudes for galaxies whenever possible, adopting them
if the associated aperture appears to collect all the light from an object
and if there is good agreement between these measurements and the ``model''
magnitudes.  For the majority of AGN hosts, the redshift-based distances and
Petrosian magnitudes are indeed suitable.  However, we identified a number
of cases where it is likely that either the distance or brightness (and thus
mass) has been underestimated, which caused us to remove several objects from
our low-mass sample. A few objects with distance or photometry issues still
qualify as dwarf galaxies when their stellar masses are revised; the
problems and their remedies are described in \S~5.5.

We corrected the $gri$ magnitudes of the remaining low-mass AGN candidates
for Galactic reddening (Schlafly \& Finkbeiner 2011) and applied
$K$-corrections using the methods of Chilingarian et al.\ (2010), which,
for this low-redshift sample, typically affect only the value of $g$, and
by at most a few hundreths of a magnitude.
Also, since the emission lines of some objects have very high equivalent
widths, we made adjustments to the galaxy magnitudes for the non-stellar
flux from the active nucleus (details are provided in \S~5.3.2).  We then
selected all AGNs whose host galaxies have stellar masses (via Bell et al.\
2003) of $10^{10}\; M_{\odot}$ or less.  The typical IMBH candidate discovered
in previous SDSS surveys (e.g., Barth et al.\ 2008) resides in a galaxy with
a stellar mass in excess of this value (see \S~5.4).

In total, we have identified 28 dwarf galaxies with accreting central black
holes.  The SDSS images of the objects and their continuum-subtracted spectra
in the H$\alpha$ and H$\beta$ regions are displayed in Figure~4.  Table~1
lists the SDSS names, distance information, photometry data (which include
the corrections discussed above), absolute $g$ magnitudes, and stellar masses
for the galaxies, along with any designations they may have in well-known
catalogs (i.e., NGC, UGC, IC, etc.).

Table 2 contains emission-line data relevant to the classifications of
the objects.  In addition to the commonly reported line-flux ratios, we
also include the \ion{He}{2} $\lambda 4686$/H$\beta$ ratio for those with
detected \ion{He}{2} emission.  The locations of our dwarf galaxy AGNs on
the diagnostic flux-ratio diagrams are shown in Figure~5.  Their positions
with respect to the ``maximum starburst'' lines of Kewley et al.\ (2006)
indicate that all, with two exceptions (see below), are clearly Seyfert
nuclei.  Only those objects with firm evidence of broad Balmer-line
emission are classified as type~1 Seyferts in Table~2; the remainder are
listed as Seyfert~2s.  The classifications of the two objects denoted as
``Sy2:'' in the Table are less certain than the others; their emission
lines are weak and the $S/N$ ratios of their raw spectra are modest.  They
pass all of our tests, but should be confirmed as Seyferts with better data.

Two objects we have classified as Seyfert~2s do not quite meet the AGN
criteria outlined above.  The first, J0948+0958, has a high $S/N$ ratio
spectrum with line ratios that place it just inside the \ion{H}{2} galaxy
region on the [\ion{N}{2}]/H$\alpha$ plot in Figure~5.  However, it is well
separated from the locus of star-forming galaxies shown, and in terms of its
[\ion{S}{2}]/H$\alpha$ and [\ion{O}{1}]/H$\alpha$ ratios, the object resides
comfortably amongst the other dwarf-galaxy AGNs in our sample.  Moreover, it
has a strong \ion{He}{2} $\lambda 4686$ line (ionization potential = 54 eV).
Depending on whether the \ion{He}{2} profile is constrained to have the same
velocity width as the Balmer lines or is premitted to be broader (which
results in a better fit), a \ion{He}{2}/H$\beta$ ratio of 0.10--0.15 is
indicated.  This is close to the values expected for AGNs (e.g., Ferland \&
Netzer 1983) and far greater than those observed for pure star-forming
galaxies in our sample, which have \ion{He}{2}/H$\beta$ ratios of $\sim
0.02$ or less.

\vskip 0.5truein
\begin{figure}[!htb]
\vskip -1.4truein
\hskip -0.60truein
\includegraphics[trim=0 240 0 0,clip,width=1.2\textwidth]{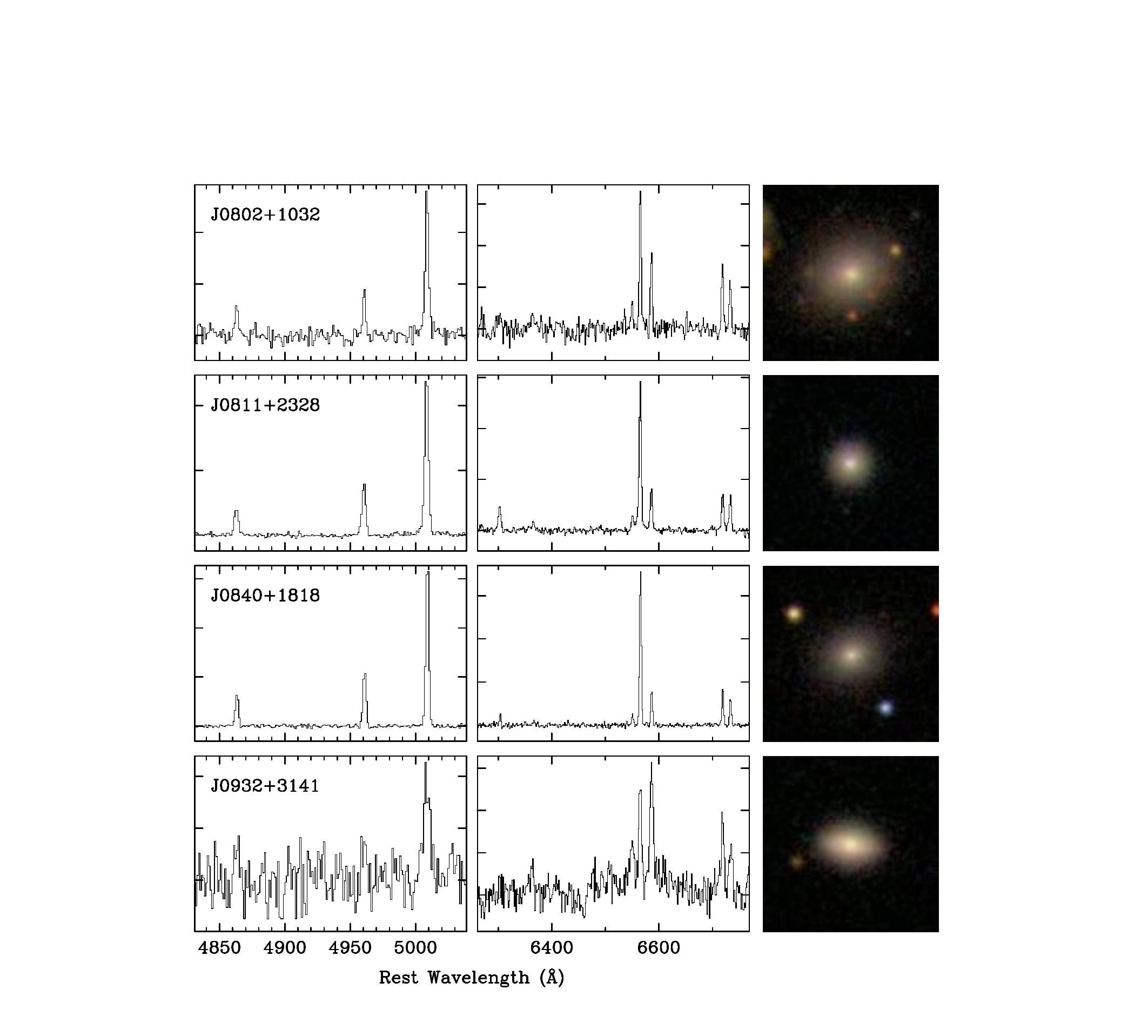}
\end{figure}
\vskip -0.2truein
{\narrower\noindent\small Fig.~4.---The dwarf-galaxy AGN sample.  Displayed
in the left and center panels are the continuum-subtracted spectra in the
H$\beta$ and H$\alpha$ regions (the latter includes the [O~{\sc i}]
$\lambda 6300$ line).  The SDSS images of the objects, which have a physical
scale of 12~kpc $\times$ 12~kpc, are shown on the right.\par}

\clearpage
\begin{figure}[!htb]
\vskip -1.4truein
\hskip -0.65truein
\includegraphics[width=1.2\textwidth]{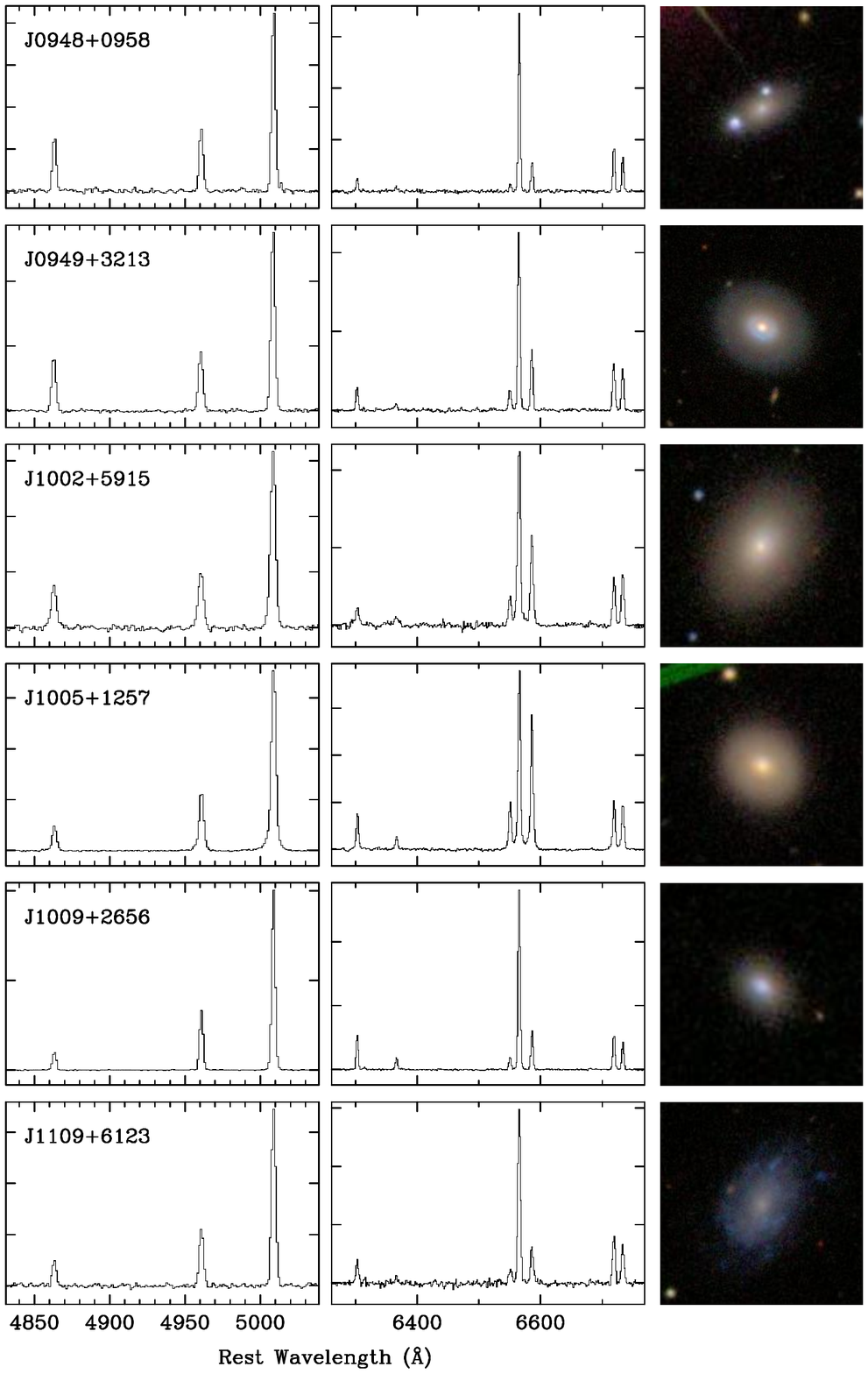}
\vskip -0.55truein
\centerline{{\small Fig.~4.---{\it continued}}}
\end{figure}

\clearpage
\begin{figure}[!htb]
\vskip -1.4truein
\hskip -0.65truein
\includegraphics[width=1.2\textwidth]{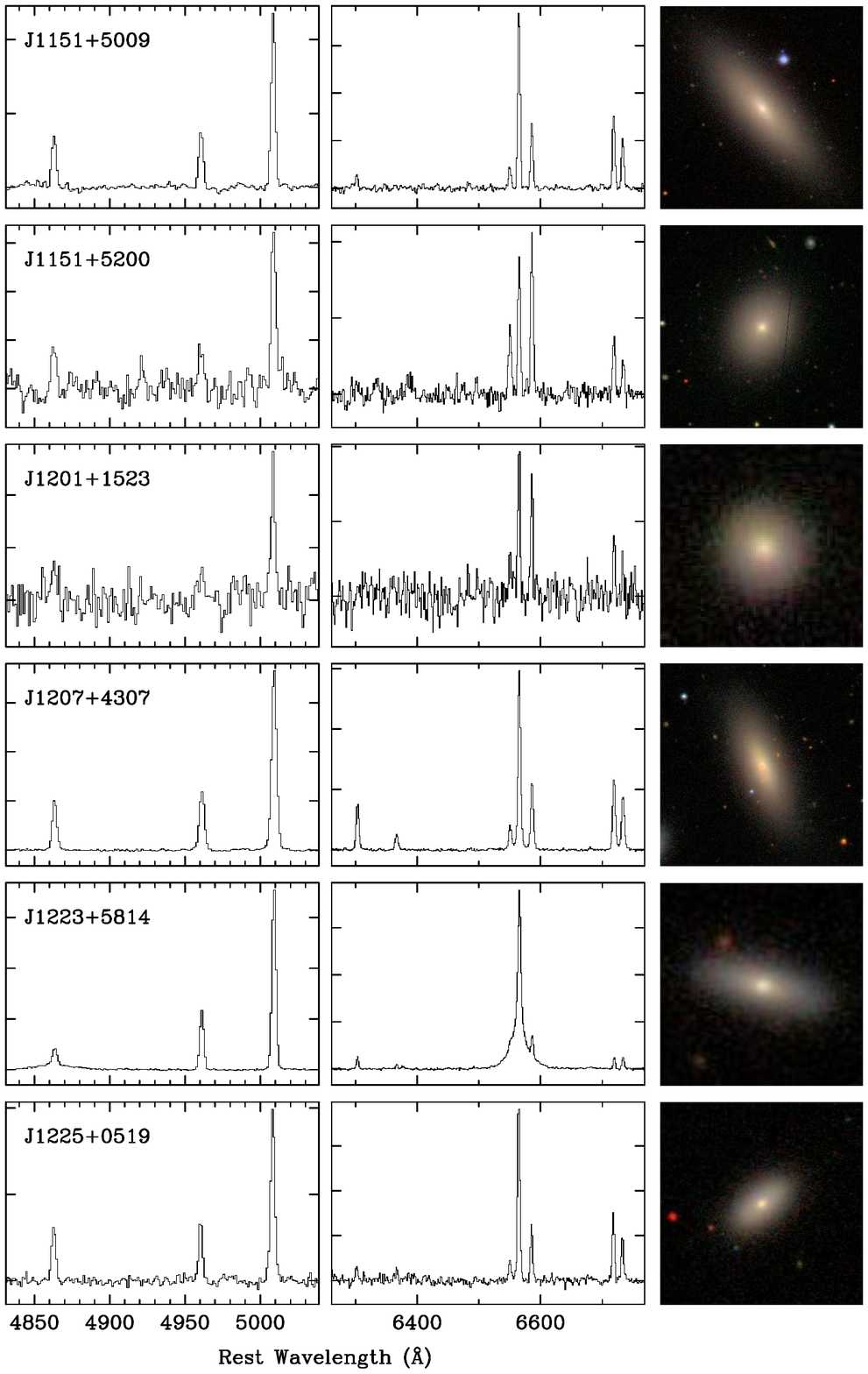}
\vskip -0.55truein
\centerline{{\small Fig.~4.---{\it continued}}}
\end{figure}

\clearpage
\begin{figure}[!htb]
\vskip -1.4truein
\hskip -0.65truein
\includegraphics[width=1.2\textwidth]{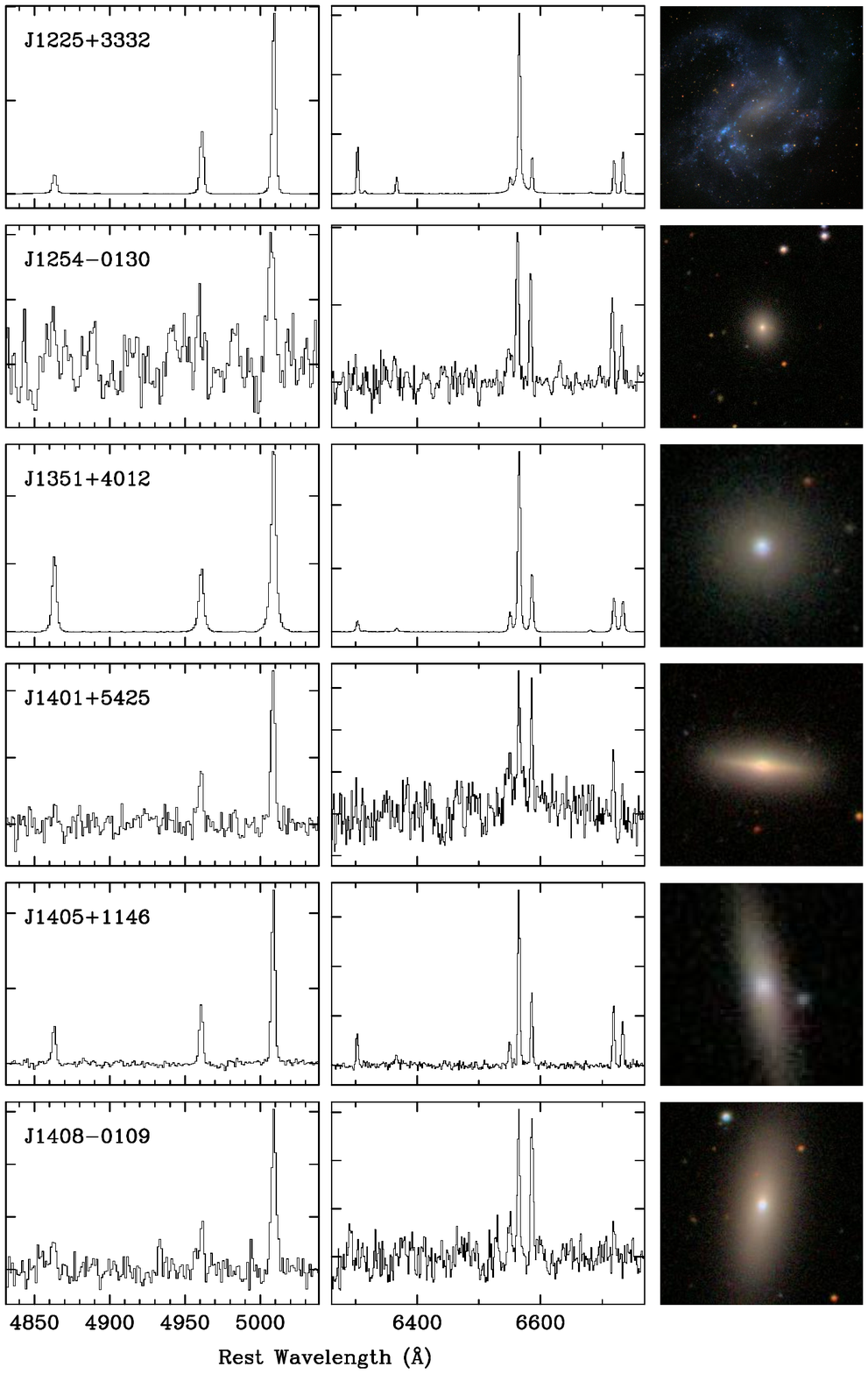}
\vskip -0.55truein
\centerline{{\small Fig.~4.---{\it continued}}}
\end{figure}

\clearpage
\begin{figure}[!htb]
\vskip -1.4truein
\hskip -0.65truein
\includegraphics[width=1.2\textwidth]{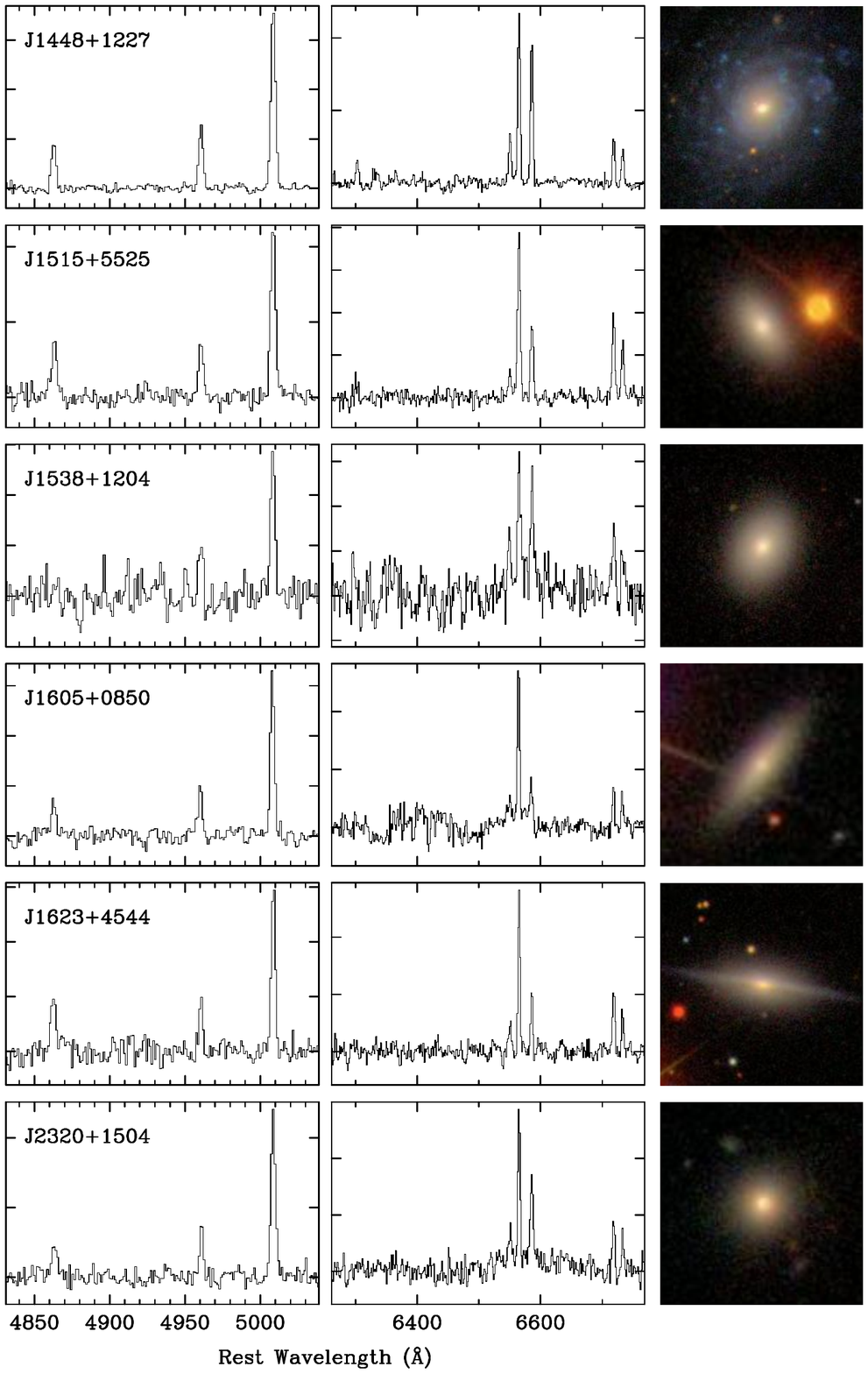}
\vskip -0.55truein
\centerline{{\small Fig.~4.---{\it continued}}}
\end{figure}

\clearpage
\begin{center}
\begin{deluxetable}{lcccccccl}
\tabletypesize{\scriptsize}
\rotate
\tablecaption{Galaxy Properties}
\tablewidth{0pt}
\tablehead{
\colhead{SDSS Name} & \colhead{$cz$ (km s$^{-1}$)} & \colhead{$d$ (Mpc)} &
\colhead{$g$ (mag)} & \colhead{$g-r$ (mag)} & \colhead{$i$ (mag)} &
\colhead{$M_g$ (mag)} & \colhead{$M_{\star}/10^9\, M_{\odot}$} &
\colhead{Catalog Names}
}
\startdata
J080212.06+103234.1                  &  4358 &	62.2 &	16.10 &	0.66 &	15.06 &	$-$17.87 &   6.5  &  \nodata           \\
J081145.29+232825.7                  &	4729 &	67.4 &	16.91 &	0.58 &	16.03 &	$-$17.23 &   2.7  &  \nodata           \\
J084025.54+181858.9                  &	4483 &	64.9 &	16.84 &	0.61 &	15.95 &	$-$17.22 &   2.8  &  \nodata           \\
J093251.11+314145.0                  & 	4595 &	67.2 &	16.07 &	0.68 &	15.04 &	$-$18.07 &   8.1  &  \nodata           \\
J094800.79+095815.4\tablenotemark{a} &  3110 &  47.3 &  16.51 & 0.48 &  15.73 & $-$16.87 &   1.4  &  \nodata           \\
J094941.20+321315.9                  & 	1537 &	26.2 &	14.10 &	0.57 &	13.24 &	$-$17.99 &   5.1  &  NGC 3011, Mrk 409 \\  
J100200.96+591508.3                  & 	2805 &	43.2 &	14.72 &	0.60 &	13.78 &	$-$18.45 &   9.1  &  MCG +10.15.001    \\
J100551.19+125740.6                  & 	2801 &	43.3 &	14.93 &	0.67 &	13.89 &	$-$18.25 &   9.4  &  IRAS F10031+1312  \\  
J100935.66+265648.9                  & 	4302 &	64.1 &	16.87 &	0.48 &	16.14 &	$-$17.16 &   1.8  &  \nodata           \\
J110912.39+612346.6                  & 	2014 &	33.3 &	15.46 &	0.27 &	15.14 &	$-$17.15 &   0.8  &  UGC 6192          \\
J115113.43+500924.8                  & 	~935 &	18.7 &	13.32 &	0.67 &	12.30 &	$-$18.03 &   7.6  &  NGC 3922          \\
J115113.45+520003.0                  & 	~905 &	18.2 &	13.87 &	0.67 &	12.86 &	$-$17.43 &   4.3  &  NGC 3931          \\
J120108.03+152353.6                  & 	5176 &	78.4 &	16.07 &	0.65 &	15.09 &	$-$18.40 &   9.9  &  \nodata           \\
J120746.11+430734.8                  & 	~934 &	18.5 &	13.61 &	0.76 &	12.47 &	$-$17.73 &   7.6  &  NGC 4117          \\
J122342.81+581446.2                  & 	4355 &	65.3 &	16.05 &	0.60 &	15.12 &	$-$18.02 &   6.0  &  \nodata           \\  
J122505.60+051944.7                  & 	2000 &	33.4 &	15.22 &	0.65 &	14.23 &	$-$17.40 &   4.0  &  \nodata           \\
J122548.86+333248.7\tablenotemark{b} & ~312  &  ~4.3 &  10.58 & 0.45 &  ~9.98 & $-$17.28 &   1.1  &  NGC 4395          \\
J125401.88$-$013030.0                &  1136 &	19.0 &	15.18 &	0.66 &	14.19 &	$-$16.22 &   1.3  &  \nodata           \\
J135125.37+401247.8                  & 	2465 &	41.3 &	15.02 &	0.58 &	14.18 &	$-$18.06 &   5.5  &  Mrk 462           \\
J140116.03+542507.5                  &	1893 &	33.0 &	14.44 &	0.72 &	13.33 &	$-$18.16 &  10.0~ &  MCG +09.23.005    \\
J140510.39+114616.9                  & 	5235 &	80.6 &	16.40 &	0.52 &	15.62 &	$-$18.13 &   5.0  &  IC 976            \\
J140843.28$-$010941.9                &	1532 &	26.0 &	13.41 &	0.55 &	12.56 &	$-$18.67 &   9.2  &  UGC 9040          \\
J144842.57+122725.9                  & 	1798 &	31.6 &	13.90 &	0.55 &	13.07 &	$-$18.60 &   8.4  &  NGC 5762          \\  
J151500.96+552555.2                  & 	3179 &	50.1 &	15.40 &	0.65 &	14.28 &	$-$18.10 &   8.4  &  \nodata           \\  
J153851.70+120450.2                  & 	1992 &	34.3 &	14.27 &	0.64 &	13.29 &	$-$18.41 &   9.7  &  IC 1131           \\
J160544.57+085043.9\tablenotemark{c} &  4661 &  72.2 &  16.66 & 0.69 &  15.47 & $-$17.63 &   6.4  &  \nodata           \\
J162335.07+454443.4                  & 	1858 &	32.6 &	14.87 &	0.67 &	13.83 &	$-$17.69 &   5.7  &  UGC 10370         \\
J232028.21+150420.9                  & 	3843 &	52.9 &	15.75 &	0.67 &	14.72 &	$-$17.87 &   6.6  &  \nodata           \\
\enddata
\tablenotetext{a}{DR7 model magnitudes are listed.}
\tablenotetext{b}{Values listed in the $g$, $g-r$, and $i$ columns for this object are its $B$, $B-V$, and $K$
magnitudes from the RC3 and 2MASS catalogs. The absolute $g$ magnitude is estimated from these using the Lupton
transformations (see \S~5.5).  The stellar mass is calculated via the $BVK$ magnitudes and the appropriate coefficients
from Bell et al.\ (2003).}
\tablenotetext{c}{DR8 Petrosian magnitudes are listed.}
\end{deluxetable}
\end{center}

\clearpage
\begin{center}
\begin{deluxetable}{lcccccccc}
\tabletypesize{\scriptsize}
\rotate
\tablecaption{Narrow Emission-Line Properties}
\tablewidth{0pt}
\tablehead{
\colhead{Object} & \colhead{Class} & \colhead{He {\sc ii}/H$\beta$} & \colhead{[O {\sc iii}]/H$\beta$} &
\colhead{[N {\sc ii}]/H$\alpha$} & \colhead{[S {\sc ii}]/H$\alpha$} & \colhead{[O {\sc i}]/H$\alpha$} &
\colhead{log $L_{5007}/L_{\odot}$\tablenotemark{a}} & \colhead{log $M_{\rm BH,min}/M_{\odot}$}
}
\startdata
J080212.06+103234.1&	Sy2  &	\nodata &	5.12 &	0.53 & 	    0.81 &	\nodata &	5.09 & 3.6\\
J081145.29+232825.7&	Sy2  &	\nodata &	6.06 &	0.24 & 	    0.44 &	   0.15 &	5.91 & 4.4\\
J084025.54+181858.9&	Sy2  &	   0.13 &	5.35 &	0.22 & 	    0.41 &	   0.07 &	5.87 & 4.3\\
J093251.11+314145.0&   ~Sy2: &	\nodata & $>$ 3.16~~~ &	1.32 & 	    1.36 &	\nodata &	4.99 & 3.5\\
J094800.79+095815.4& 	Sy2  &	   0.10 &	3.42 &	0.16 & 	    0.44 &	   0.07 &	5.73 & 4.2\\
J094941.20+321315.9& 	Sy2  &	   0.17 &	3.47 &	0.36 & 	    0.53 &	   0.13 &	5.66 & 4.1\\
J100200.96+591508.3& 	Sy2  &	\nodata &	4.14 &	0.41 & 	    0.46 &	   0.13 &	5.63 & 4.1\\
J100551.19+125740.6& 	Sy2  &	   0.21 &	8.49 &	0.76 & 	    0.51 &	   0.19 &	6.29 & 4.8\\
J100935.66+265648.9& 	Sy2  &	   0.27 &	9.15 &	0.22 & 	    0.34 &	   0.19 &	6.63 & 5.1\\
J110912.39+612346.6& 	Sy2  &	   0.23 &	7.15 &	0.21 & 	    0.49 &	   0.12 &	5.34 & 3.8\\
J115113.43+500924.8& 	Sy2  &	   0.14 &	3.17 &	0.36 & 	    0.69 &	   0.06 &	4.99 & 3.5\\
J115113.45+520003.0& 	Sy2  &	\nodata &	3.92 &	1.24 & 	    0.72 &	\nodata &	4.16 & 2.6\\
J120108.03+152353.6& 	Sy2  &	\nodata &	3.42 &	0.78 & $>$ 0.68~~~~&	\nodata &	4.98 & 3.4\\
J120746.11+430734.8& 	Sy2  &	   0.05 &	3.11 &	0.39 &	    0.68 &	   0.25 &	5.56 & 4.0\\
J122342.81+581446.2& 	Sy1  &	   0.08 &	9.23 &	0.19 &	    0.18 &	   0.09 &	6.53 & 5.0\\
J122505.60+051944.7& 	Sy2  &	   0.19 &	3.15 &	0.29 &	    0.59 &	   0.08 &	5.12 & 3.6\\
J122548.86+333248.7& 	Sy1  &	   0.24 &	9.76 &	0.22 &	    0.48 &	   0.30 &	5.31 & 3.8\\
J125401.88$-$013030.0&   ~Sy2: &	\nodata & $>$ 3.12~~~ &	0.65 &	    0.92 &	\nodata &	3.81 & 2.3\\
J135125.37+401247.8& 	Sy2  &	   0.08 &	2.92 &	0.32 &	    0.36 &	   0.06 &	6.54 & 5.0\\
J140116.03+542507.5&	Sy2  &	\nodata & $>$ 7.30~~~ &	0.87 & $>$ 0.92~~~~&	\nodata &	4.74 & 3.2\\
J140510.39+114616.9& 	Sy2  &	   0.16 &	4.55 &	0.43 &	    0.60 &	   0.18 &	5.90 & 4.4\\
J140843.28$-$010941.9&	Sy2  &	\nodata &	5.28 &	1.06 &	$>$ 0.37~~~~&	\nodata &	4.75 & 3.2\\
J144842.57+122725.9& 	Sy2  &	\nodata &	4.12 &	0.85 &	    0.47 &	   0.14 &	5.39 & 3.9\\
J151500.96+552555.2& 	Sy2  &	   0.26 &	3.02 &	0.45 &	    0.82 &	   0.12 &	5.20 & 3.7\\
J153851.70+120450.2& 	Sy2  &	\nodata & $>$ 5.00~~~ &	0.79 & $>$ 1.03~~~~&	\nodata &	4.65 & 3.1\\
J160544.57+085043.9& 	Sy2  &	\nodata &	4.20 &	0.35 &	    0.44 &	\nodata &	5.33 & 3.8\\
J162335.07+454443.4& 	Sy2  &	\nodata &	3.03 &	0.38 &	    0.67 &	\nodata &	4.55 & 3.0\\
J232028.21+150420.9& 	Sy2  &	\nodata &	4.69 &	0.61 &	    0.60 &	\nodata &	5.25 & 3.7\\
\enddata
\tablenotetext{a}{$L_{\odot} = 3.8 \times 10^{33}$ ergs~s$^{-1}$.}
\end{deluxetable}
\end{center}

\clearpage
\begin{figure}[!htb]
\vskip -0.30truein
\hskip -0.27truein
\includegraphics[trim=0 150 0 150,clip,width=1.05\textwidth]{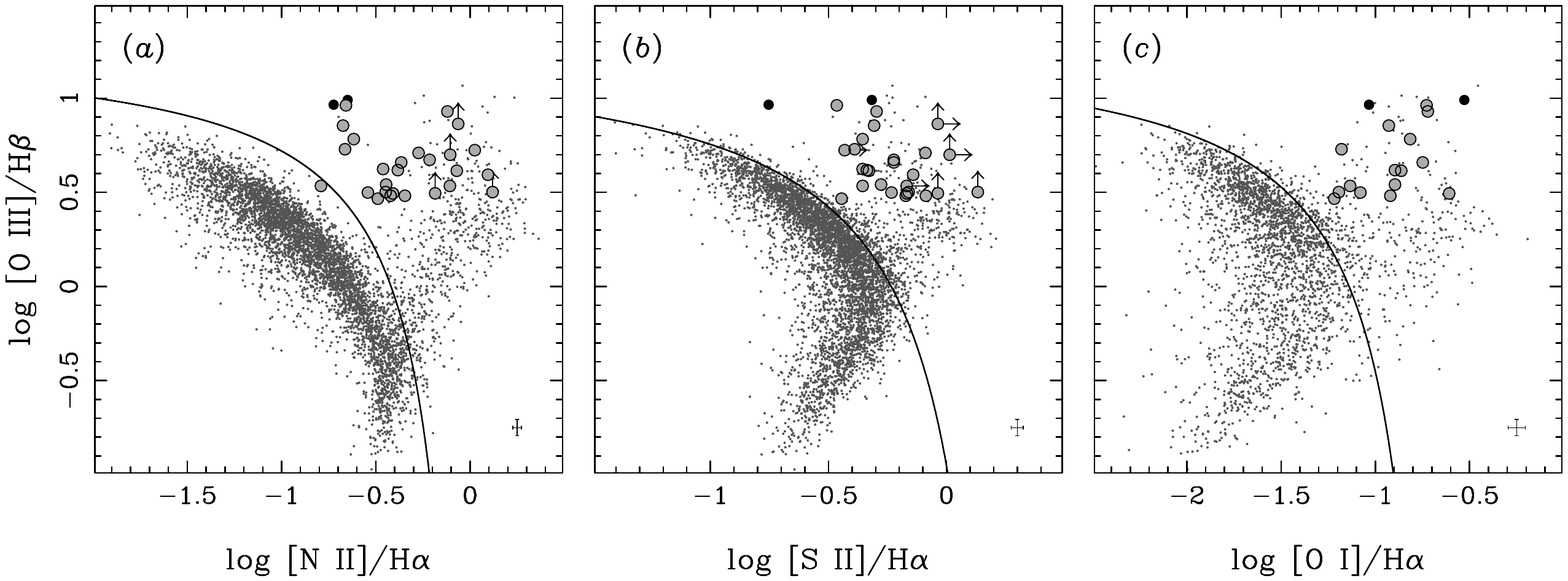}
\end{figure}
{\narrower\noindent\small Fig.~5.---Locations of the dwarf-galaxy AGNs in our
sample (circles) on standard emission-line diagnostic diagrams.  Solid symbols
represent the two objects with clear detections of broad emission lines.
Lower limits on the [\ion{O}{3}]/H$\beta$ and [\ion{S}{2}]/H$\alpha$ ratios
are indicated with arrows.  Typical error bars, computed for the object in
the sample that has the median $S/N$ ratio for all detected diagnostic lines,
are shown in the lower-right corner of each panel.  The grey dots represent
all SDSS DR7 galaxies within 80 Mpc that have detections of the relevant
emission lines (5345 in the [\ion{N}{2}] and [\ion{S}{2}] plots, and 2826
in the [\ion{O}{1}] plot).  Empirical ``maximum starburst'' lines (Kewley
et al.\ 2006) separate the AGN and \ion{H}{2} galaxy regions on the plots.\par}
\vskip 0.15truein

The second object, J1351+4012, lies in the AGN regions on all of the plots
in Figure~5, but it has relatively weak low-ionization forbidden lines and
[\ion{O}{3}]/H$\beta$ $< 3$ (although, because the [\ion{O}{3}] lines are
broader than H$\beta$, the ratio is closer to 3 than the spectrum in Fig.~4
might suggest).  Similar to J0948+0958, this object also has a fairly strong
\ion{He}{2} line with \ion{He}{2}/H$\beta$ = 0.08.  Weak [\ion{Fe}{7}]
$\lambda 6087$ and [\ion{Fe}{10}] $\lambda 6375$ lines, both of which have
ionization potentials in excess of of 100 eV, appear to be present as well.
The spectra of J0948+0958 and J1351+4012 near \ion{He}{2} $\lambda 4686$
are displayed in Figure~6.

The combination of strong \ion{He}{2} emission and the fact that some of
the flux ratios involving low-ionization lines are consistent with those of
Seyfert nuclei suggests that these galaxies are powered in part by AGNs.
Without the benefit of additional data, our best hypothesis is that they
are composite objects whose SDSS spectra contain emission from both an
active nucleus and circumnuclear star-forming regions.  Relative to a pure
AGN spectrum, we would expect either the H$\alpha$ or H$\beta$ emission
line to be enhanced in this scenario, which would explain the ambiguous
line ratios of the objects.  Indeed, the central region of J1351+4012 is
very blue (its ``fiber'' magnitudes measured in the same area covered by
the spectroscopic fibers indicate $g-r = 0.39$) and its continuum spectrum
exhibits strong Balmer absorption, confirming the presence of a young
stellar population.
\negthinspace\negthinspace\negthinspace
\footnote{J1351+4012 would also qualify as ``composite'' according
to Kewley et al.\ (2006), owing to the fact that it falls between the
empirical and theoretical (Kewley et al.\ 2001) maxiumum starburst lines
on the [\ion{O}{3}]/H$\beta$ vs.\ [\ion{N}{2}]/H$\alpha$ diagram.  We note,
however, that the one other object in our sample that would be composite by
this definition, J1225+0519, displays no evidence of ongoing or recent star
formation.  It is very red (in terms of both its total and fiber magnitudes)
and its nuclear spectrum is dominated by an old stellar population.}
Similarly, J0948+0958 is one of our bluer objects ($g-r = 0.48$).
Being low-mass galaxies, J0948+0958 and J1351+4012 are valuable additions
to our AGN sample.  They also suggest new ways of identifying active black
holes in star-forming dwarf galaxies.

\vskip 0.15truein
\begin{figure}[!htb]
\vskip -0.30truein
\hskip 1.15truein
\includegraphics[trim=0 50 0 200,clip,width=1.0\textwidth]{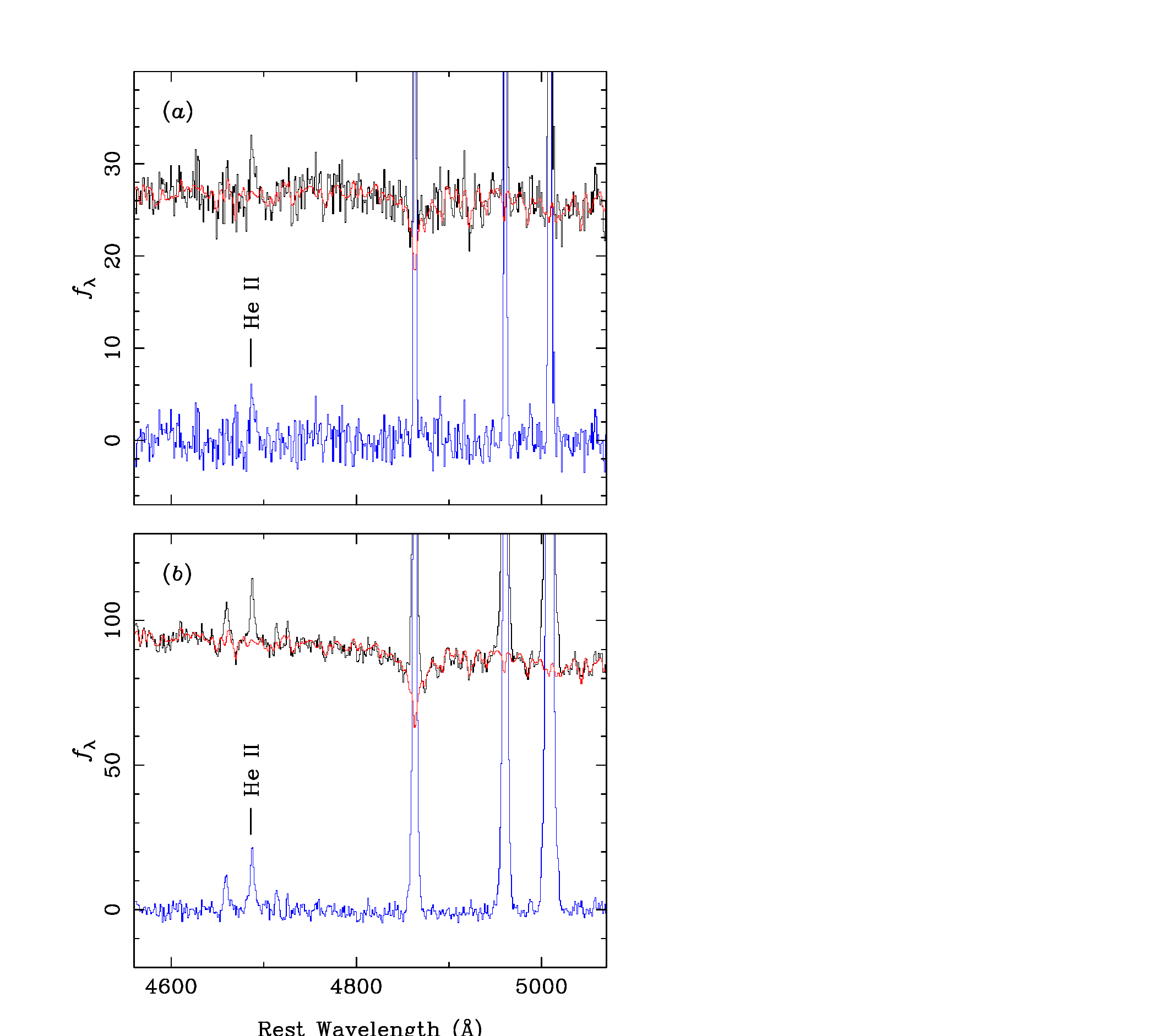}
\end{figure}
\vskip 0.05truein
{\narrower\noindent\small Fig.~6.---SDSS spectrum (black), continuum fit (red),
and emission-line residuals (blue) in the vicinity of \ion{He}{2}
$\lambda 4686$ for ($a$) J0948+0958 and ($b$) J1351+4012, the two objects
in our sample with ambiguous classifications.  Although the narrow-line flux
ratios of these objects do not strictly satisfy the Seyfert criteria we have
adopted, the strength of the \ion{He}{2} line (relative to H$\beta$) in each
is typical of that observed in AGNs.  In J1351+4012, \ion{He}{2} is flanked
by [\ion{Ar}{4}] $\lambda 4712$ to the red and what appears to be
[\ion{Fe}{3}] $\lambda 4658$ to the blue.\par}

\subsection{Nuclear Properties}

The nuclear emission-line characteristics of the galaxies in our sample have
played a primary role in their identification as AGNs.  Bearing in mind that
the continuum-subtracted spectra have a wide range of line strengths and
$S/N$ ratios, we have examined these characteristics more closely to gain
additional insight into the nature of the objects.

\subsubsection{Narrow Emission Lines}

The narrow emission lines of AGNs contain information about the continuum
source and the properties of the associated photoionized gas.  Following
Ludwig et al.\ (2012), who examined a subset of type~1 IMBH candidates from
the SDSS surveys of Greene \& Ho (2004, 2007b), we have measured the narrow
\ion{He}{2} $\lambda 4686$/H$\beta$ flux ratios for our objects when possible.
This ratio provides an estimate of the slope of the ionizing continuum in
the far-ultraviolet (Penston \& Fosbury 1978) and an opportunity to compare
the AGNs uncovered in our survey with existing samples of IMBH candidates.
As Table~2 indicates, \ion{He}{2}/H$\beta$ = 0.05--0.27 for the 14 objects
with a detected \ion{He}{2} line.  Assuming the UV continuum between 228~\AA\
and 912~\AA\ can be described as a power law (i.e., $F_{\nu} \propto
\nu^{-\alpha_{\rm UV}}$), this translates to a spectral index range of
$\alpha_{\rm UV}$ = 1.5--2.6, with a median value of 1.8.  This distribution
of power-law slopes is very similar to that obtained by Ludwig et al.\ (2012),
suggesting that there are no significant differences in the ionizing continuua
of their more luminous broad-line objects and the predominantly narrow-line
AGNs in our sample.

Weak [\ion{O}{3}] $\lambda 4363$ emission is detected in the spectra of a
dozen members of our sample.  The measured fluxes, along with those of
the other [\ion{O}{3}] lines, indicate temperature-sensitive
$F(\lambda 4959 + \lambda 5007) / F(\lambda 4363)$ ratios of $\sim$ 40--80,
which correspond to electron temperatures of $T_e \approx (1.4 - 2.0) \times
10^4$~K in the narrow-line region (NLR) gas (Osterbrock 1989).  Although
some of the objects lacking $\lambda 4363$ detections could have lower
temperatures (say, $\sim 10^4$~K), the median value for those with detections
($T_e \approx 1.5 \times 10^4$~K, based on a flux ratio of 63) is likely to
be representative of the sample as a whole.

Assuming the [\ion{O}{3}] and [\ion{S}{2}] lines are produced in the same
regions,
\negthinspace\negthinspace
\footnote{This may not be a safe assumption if the NLRs of our objects
have significant density or ionization stratification, which is observed in
some (but not all) luminous AGNs (e.g., Filippenko \& Halpern 1984; Veilleux
1991).}
we can estimate the electron densities in the NLRs of our objects by
combining this temperature and the density-sensitive
[\ion{S}{2}] $F(\lambda 6716)/F(\lambda 6731)$ flux ratio.
Most of the AGNs in our sample have [\ion{S}{2}] ratios that are tightly
clustered around a value of $\sim 1.4$, indicating NLR densities near the
low-density limit.  In fact, 20/28 objects have
$F(\lambda 6716)/F(\lambda 6731) \ge 1.2$, implying $n_e \le 10^2$~cm$^{-3}$
(Osterbrock 1989).  The electron densities of the remaining 8 objects, with
[\ion{S}{2}] ratios between 0.8 and 1.1, are in the $\sim$ 500--1200~cm$^{-3}$
range. 

One of the most striking aspects of the spectra shown in Figure~4 is that
some objects exhibit exceptionally low [\ion{N}{2}]/H$\alpha$ ratios, with
values in the vicinity of $\sim 0.2$.  As Table~2 indicates, about half of
our galaxies have [\ion{N}{2}]/H$\alpha$ $\le 0.4$ --- lower than the ratios
observed for the vast majority of objects classified as AGNs in large SDSS
surveys (e.g., Kewley et al.\ 2006).  Weak [\ion{N}{2}] Seyferts, which are
also present in the IMBH samples of Barth et al.\ (2008) and Ludwig et al.\
(2012), are good candidates for objects that have lower metal abundances in
their NLR gas compared to classical Seyfert nuclei (Groves et al.\ 2006).
As the latter reside in massive galaxies, this might be expected based on
the observed mass-metallicity relation in galaxies (e.g., Tremonti et al.\
2004).  However, the ionization parameter in the NLR --- i.e., the ratio
of the densities of ionizing photons and gas --- affects the [\ion{N}{2}]
strength as well (e.g., Ferland \& Netzer 1983), implying that the density
and geometry of the NLR are also important factors.

We have used the photoionzation calculations performed by Ludwig et al.\
(2012, based on the models of Groves et al.\ 2004) to explore the effects
of metallicity and ionization parameter in our sample, particularly for
those objects with the lowest [\ion{N}{2}]/H$\alpha$ ratios.  In their
Figure~6, Ludwig et al.\ (2012) present predictions of the
[\ion{O}{3}]/H$\beta$ and [\ion{N}{2}]~$\lambda\lambda$6548,6583/H$\alpha$
flux ratios for a range of metallicity and ionization parameter values.  For
a continuum slope of $\alpha_{\rm UV} = 1.7$ and a density of $n_e = 100$
cm$^{-3}$ or $n_e = 1000$ cm$^{-3}$ --- appropriate assumptions for our
sample --- the Ludwig et al.\ calculations indicate that 6/7 of our objects
with [\ion{N}{2}]/H$\alpha$ $\approx 0.2$ are located in a region of high
ionization parameter ($\log U \approx -1$) and solar (or slightly super-solar)
metallicity.  As discussed by Ludwig et al., the determination of absolute
metallicities is difficult, but combining all of the available spectral
evidence, our findings suggest that a high ionization parameter is mainly
responsible for the weakness of [\ion{N}{2}] in these objects.  We note,
however, that the ambiguous line ratios of the seventh weak-[\ion{N}{2}]
object, J0948+0958, might be explained (in part) by NLR gas with sub-solar
abundances.  We speculated above in \S~5.1 that the SDSS spectrum of
J0948+0958 may be a combination of emission from an AGN and cirumnuclear
\ion{H}{2} regions, but the [\ion{O}{3}]/H$\beta$ and [\ion{N}{2}]/H$\alpha$
ratios of this galaxy are consistent with the $Z = 0.5\, Z_{\odot}$ model
calculations in the $\alpha_{\rm UV} = 1.7$, $n_e = 100$ cm$^{-3}$ plot shown
in Figure~6 of Ludwig et al.  Interestingly, AGN-like [\ion{S}{2}]/H$\alpha$,
[\ion{O}{1}]/H$\alpha$, and \ion{He}{2}/H$\beta$ flux ratios, which
J0948+0958 displays, are also expected in this scenario (Groves et al.\
2006).  Spatially resolved spectra of this object would help clarify its
nature; in any case, it represents the best candidate for a low-metallicity
AGN in our sample.

\subsubsection{Broad Emission Lines}

Examination of Figure~4 reveals that there is a paucity of objects with
obvious broad emission lines in our sample.  In fact, there are just two:\
one is J1225+3332, which is NGC~4395 itself, and the other is J1223+5814,
the only object in the Greene \& Ho (2007b) IMBH survey that falls within
the redshift and stellar-mass limits of our study.  The broad H$\alpha$
components in both objects are fairly narrow ($< 1000$ km~s$^{-1}$ FWHM),
and similar to the broad lines in other type~1 IMBH candidates (e.g.,
Greene \& Ho 2004), their profiles have extended, Lorentzian-like wings.

Figure~4 indicates that two other AGNs in our sample, J0932+3141 and
J2320+1504, have features in their residual spectra reminiscent of
broad H$\alpha$ lines.  Indeed, a reanalysis of the spectra with GANDALF
results in an improved fit when a broad-line component is included.
However, the putative broad-line signal in each is extremely weak
in comparison to the flux-density level in the raw data and the noise
in the residual spectrum.  In both cases, the formal $S/N$ ratio of
the fitted broad lines is only $\sim 2$.  Two other galaxies, J1401+5425
and J1538+1204, may also have broad H$\alpha$ lines, but if so, the
significance of the detections is even lower.  Until better data indicate
otherwise, we consider all four of these objects to be narrow-line AGNs.

The ``noise'' spectra (described above) of several AGNs in our sample exhibit
significant positive residuals in the wings of H$\alpha$, which might point
to the presence of a broad-line component.  However, in {\it all\/} such
cases, similar residuals are also associated with the strong forbidden lines
in the spectrum (e.g., [\ion{O}{3}]).  Modeling the emission lines using
non-Gaussian profiles (e.g., Voigt) with a single velocity width, we obtain
good fits to the spectra without the need for extra broad Balmer components.

Taken at face value, the data in hand would suggest that broad-line AGNs
are quite uncommon in local dwarf galaxies.  Only 2/28 objects in our sample
have clear evidence of broad emission lines, which is far less than one would
expect based on the 1:3 or 1:4 ratios of type~1 to type~2 Seyferts obtained
in previous optical surveys of luminous AGNs (e.g., Osterbrock \& Shaw 1988;
Maiolino \& Rieke 1995).  In terms of possible selection effects, we point
out that the detection of faint, moderately broad Balmer lines associated
with low-mass black holes could be challenging.  To draw firm conclusions
about the relative numbers of type~1 and type~2 AGNs in our sample, we would
require spectra of uniformly high $S/N$ ratio (and higher spectral resolution).
Alternatively, if the unified AGN model (e.g., Antonucci 1993) is applicable
to Seyfert nuclei in dwarf galaxies, we might expect our sample to be biased
in favor of unobscured (and hence brighter) broad-line objects, and not the
weak type~2 objects that dominate it.  The apparently low type~1/type~2 ratio
we observe could thus be a hint that the broad-line regions of AGNs in the
least luminous, least massive systems are more likely to be obscured than
those of classical Seyfert nuclei.  This is an important issue that should
be explored further.

\subsubsection{Black Hole Masses}

The kinematic information contained in broad emission lines provides the
opportunity to estimate black hole masses in AGNs (Xiao et al.\ 2011, and
references therein).  Obviously, with such a small number of type~1 AGNs,
there is little we can say about the black-hole masses of our objects.
We can, however, use Eddington-luminosity arguments to estimate the minimum
masses of their black holes (e.g., Barth et al.\ 2008).  To do this, we
take $L_{5007}$, the luminosity of the strong (and presumably isotropic)
[\ion{O}{3}] $\lambda 5007$ emission line, and compute the bolometric
luminosity $L_{\rm bol}$ by applying a bolometric correction (e.g., Heckman
et al.\ 2004).  The minimum black-hole mass is obtained by assuming the
source is radiating at its Eddington limit --- i.e., that $L_{\rm bol} =
L_{\rm Edd}$ and $L_{\rm Edd} = 1.3 \times 10^{38}\, M_{\rm BH}/M_{\odot}$
ergs~s$^{-1}$.

Included in Table~2 are the values of $L_{5007}$ for the objects in our
sample, corrected for Galactic extinction.  While it is common to adopt
the Heckman et al.\ (2004) ratio of $L_{\rm bol}/L_{5007} \approx 3500$ for
$L_{\rm bol}$ estimates, we have instead derived a bolometric correction
using the sample of broad-line AGNs from the Greene \& Ho (2007b) IMBH
survey.  These are closer analogs to the objects in our sample, and
in addition to $L_{5007}$ measurements, they have published values of
$M_{\rm BH}$ and $L_{\rm bol}/L_{\rm Edd}$ that are independent of $L_{5007}$.
In Figure~7a, we plot the distribution of $L_{\rm bol}/L_{5007}$ ratios for
174 AGNs in the main sample of Greene \& Ho.  Ignoring the few outliers
at the very low end, the distribution has a median of $\log (L_{\rm bol}/L_{5007})
= 3.0$ and an RMS scatter of 0.3 dex.

Minimum black-hole masses based on this bolometric correction for our objects
are also listed in Table~2.  The distribution, shown in Figure~7b, has a
median of $M_{\rm BH,min} =  6 \times 10^3$~$M_{\odot}$ and a scatter of a
factor of 5.  The highest values are $\sim 1 \times 10^5$~$M_{\odot}$.

\vskip -1.35truein
\begin{figure}[!htb]
\hskip -0.2truein
\includegraphics[width=0.55\textwidth]{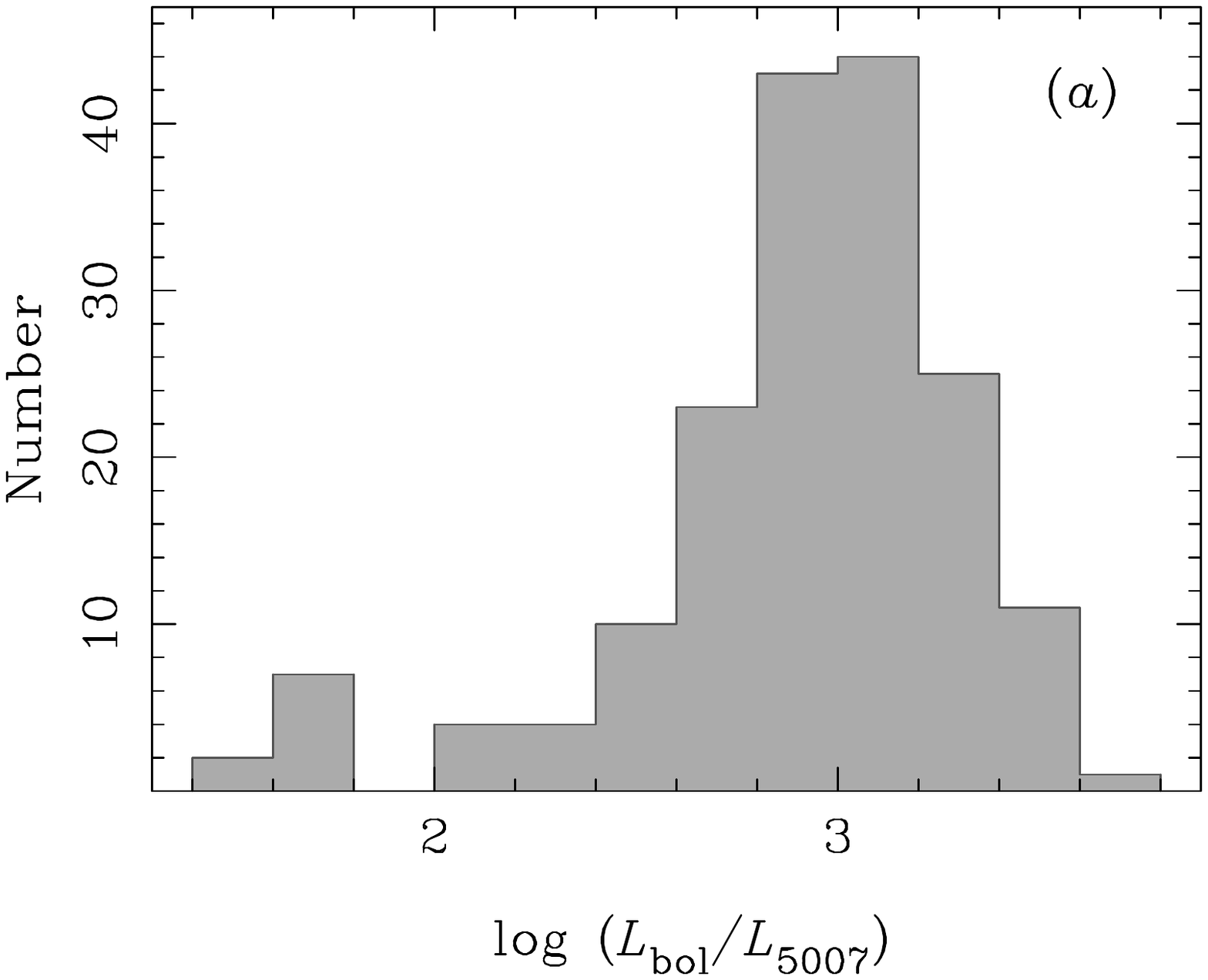}
\vskip -4.645truein
\hskip 3.0truein
\includegraphics[width=0.55\textwidth]{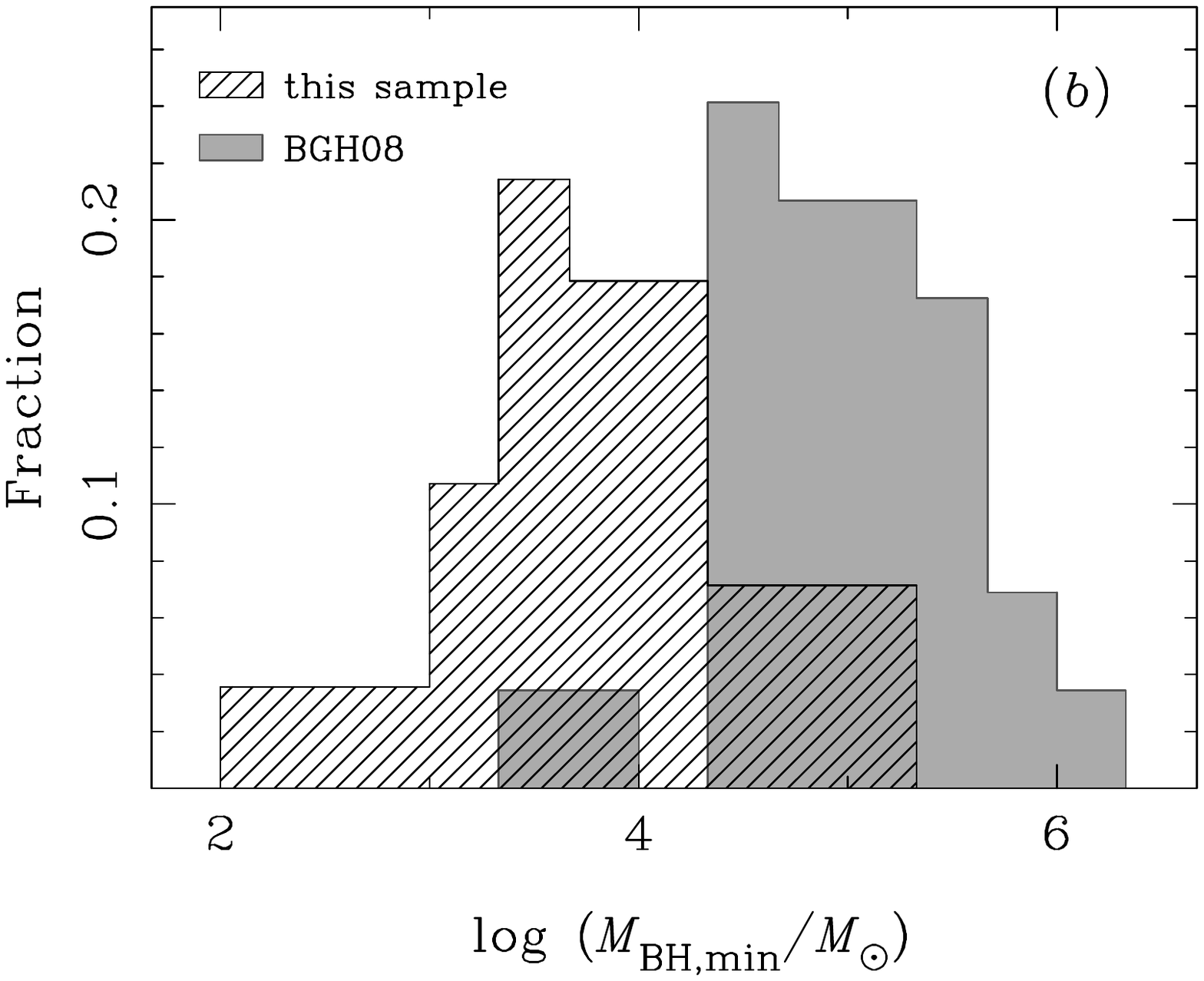}
\end{figure}

\vskip -1.0truein
{\narrower\noindent\small Fig.~7.---($a$) Distribution of
$\log (L_{\rm bol}/L_{5007})$ values for the type~1 IMBH
candidates in the main sample of Greene \& Ho (2007b).  Excluding the
handful of outliers with extremely low ratios, the distribution has a median
of 2.97 and an RMS scatter of 0.30 dex.
($b$) Distribution of minimum black-hole masses of the objects in our sample,
derived using their measured $L_{5007}$ values and the bolometric correction
implied by the $L_{\rm bol}/L_{5007}$ distribution shown in ($a$).  For
comparison, we have included the minimum black-hole masses for objects in
the IMBH sample of Barth et al.\ (2008, labeled ``BGH08''), which is similar
to ours in many respects.  As described in \S~5.4, we have taken care to
eliminte systematic differences in the approach used to make the $M_{\rm BH,min}$
estimates for the two samples.\par}
\vskip 0.02truein

The two type~1 AGNs in our sample, J1223+5814 and NGC~4395, have published
black-hole masses of $\log (M_{\rm BH}/M_{\odot}$) = 6.05 and 5.56, respectively,
based on the width and luminosity of broad H$\alpha$ (for J1223+5814; Xiao
et al.\ 2011) or the reverberation mapping technique (for NGC~4395; Peterson
et al.\ 2005).  The minimum black-hole masses we have computed imply
reasonable Eddington ratios of $\log (L_{\rm bol}/L_{\rm Edd}) = -1.1$ for
J1223+5814 and $\log (L_{\rm bol}/L_{\rm Edd}) = -1.8$ for NGC~4395.  The value
for NGC~4395 is consistent with the recent $L_{\rm bol}/L_{\rm Edd}$ estimate
by Nardini \& Risaliti (2011).

\subsubsection{Previous AGN Identifications and Non-Optical Detections}

Despite their proximity, only four of the objects in our sample are previously
classified AGNs.  In addition to NGC~4395 and J1223+5814, our sample includes
J1207+4307 (= NGC~4117), which has been part of numerous surveys of Seyfert
galaxies over the years (e.g., Ulvestad \& Wilson 1989; Cardamone et al.\
2007), but it has not been recognized as an IMBH candidate in a low-mass host
galaxy.  The other known AGN is J1109+6123, a low-mass spiral galaxy that was
included in the IMBH survey of Barth et al.\ (2008) and examined in detail by
Thornton et al.\ (2009).  All remaining 24 objects in our sample are newly
classified AGNs, and many of these --- other than having ``galaxy''
designations in the SDSS or the Two Micron All-Sky Survey (2MASS; Skrutskie
et al.\ 2006) --- are anonymous.  Among the half-dozen ``named'' objects, a
few have been classified in the literature as \ion{H}{2} galaxies; they are
discussed further in \S~5.5.

It is also interesting to note that very few of the objects in our sample
have been detected via emission that can be associated with black-hole
accretion in bands other than the optical.  Both NGC 4395 and NGC 4117 ---
two of the nearest objects --- have been studied at X-ray wavelengths (e.g.,
Moran et al.\ 1999; Cardamone et al.\ 2007), and J1223+5814 was detected in
the {\sl ROSAT\/} All-Sky Survey (Mickaelian et al.\ 2006).  One other object,
J1515+5525, was marginally detected as a serendipitous source in a 27~ks
observation with the ACIS-S instrument on board {\it Chandra}.  We have
just carried out {\it Chandra\/} observations of six additional AGNs in
our sample and can report that very few counts were detected for any of
them, even though the exposures were fairly deep (full details will be
provided elsewhere).  The situation in the radio band is similar.  A search
of the {\sl FIRST\/} survey catalog (Becker et al.\ 1995) reveals that
only three of the 27 objects observed are detected at 20~cm:\ NGC 4117
(1.9 mJy), NGC 4395 (1.2 mJy), and J1005+1257 (8.6 mJy).

The general weakness of our sample at radio and X-ray wavelengths is
important in the context of how we construct samples of IMBH candidates.
Recently, X-ray observations, either alone (e.g., Gallo et al.\ 2008;
Desroches \& Ho 2009; McAlpine et al.\ 2011; Secrest et al.\ 2012;
Araya Salvo et al.\ 2012; Schramm et al.\ 2013) or in combination with
radio observations (Reines et al.\ 2011), have been used to identify a
small number of IMBH candidates.  While it remains to be seen if
accreting IMBHs best express themselves as emission-line objects or as
sources in the X-ray and radio bands, the size and properties of our
sample suggest that, at least locally, optical techniques may provide
the most economical means for discovering low-luminosity (and potentially
the most typical) examples.

\subsection{Host Galaxy Properties}

\subsubsection{Sizes, Morphologies, and Environments}
The broadband SDSS images of the AGN host galaxies in our sample are shown
at a common scale of 12~kpc $\times$ 12~kpc in Figure~4, demonstrating that
the objects are indeed physically small.  Several of the galaxies have total
diameters of less than 6~kpc.  Figure~4 also indicates that all of the
galaxies have regular morphologies with bright, compact knots at their
centers.  But with the exception of NGC~4395 (see \S~5.5), the
continuum emission in the spectra of our objects is dominated by starlight,
implying that most of the light from these knots is associated with the
galaxies' central stellar concentrations, not their active nuclei.
Interestingly, the images also reveal that only
3/28 galaxies are obvious spirals with blue, star-forming disks.  Most of
the rest have a smooth, featureless appearance (and often reddish colors)
outside the nucleus, which might suggest that they are early-type galaxies.
However, many of the latter objects have high axis ratios (e.g., J1151+5009,
J1207+4307, J1223+5814, J1225+0519, J1405+1146, J1408$-$0109, and J1605+0850)
and must be inclined disks rather than elliptical galaxies.  Those with axis
ratios close to unity could be more face-on versions of these galaxies.  At
least one object, though --- J1623+4544 --- appears to be an edge-on S0
galaxy with a significant spheroidal component.  A detailed
characterization of the central regions of the objects in our sample would
provide valuable information about the stellar environments of their black
holes (masses, luminosities, and surface brightnesss profiles; see Greene
et al.\ 2008). Unfortunately, given the distances to some of the galaxies
and the typical seeing in SDSS images, higher resolution images are needed
for this task.

Combining a visual inspection of the SDSS images with tools available from
NED, we find that most of the objects in our sample are isolated.  None of
them is a close binary companion of a more massive galaxy, and with the
exception of two objects, none has a companion of any sort that is at roughly
the same redshift and within a projected distance of $\sim 10$ galaxy
diameters ($D_{\rm gal}$).  One of the objects with a close companion is
J1207+4307 (= NGC~4117), which lies within $\sim 1\, D_{\rm gal}$ of the dwarf
star-forming galaxy NGC~4118 and $\sim 4.5\, D_{\rm gal}$ from the LINER
NGC~4111.  The other is J1223+5814, which is separated by $\sim 3\, D_{\rm gal}$
from a fainter galaxy at about the same redshift.  Four other galaxies in our
sample are found 10--12 $D_{\rm gal}$ from galaxies at similar infall-corrected
distances:\ J1109+6123 (near NGC~3543), J1201+1504, J1515+5525 (near the
Seyfert~1 NGC~5905), and J1538+1204 (near the LINER NGC~5970).  The rest
(21/28) appear to be field galaxies.  Thus, for the majority of our sample,
clear evidence for external triggers of their nuclear activity is lacking.

\subsubsection{Colors, Stellar Populations, and Stellar Masses}

As mentioned in \S~5.1, we corrected the total magnitudes of the galaxies
in our sample for any non-stellar emission associated with their active
nuclei.  To do this, we multiplied both the original spectrum and the
residual spectrum by the SDSS filter transmission functions to determine
the contribution of the emission lines to the $gri$ fiber magnitudes.
\negthinspace\negthinspace\negthinspace
\footnote{The spectral modeling results described in \S~4 indicate that
the continua of all but one of the objects in our dwarf galaxy AGN sample
are dominated by starlight, so only the emission-line fluxes were accounted
for in this analysis.  The continuum of the one object with a significant
power-law component, J1225+3332, appears to have {\it no\/} starlight in
its spectrum, so we assigned all of the flux corresponding to the fiber
magnitudes to the AGN in this case.}
The AGN-corrected fiber magnitudes were then used to adjust the total
magnitudes of each galaxy.  In the end, this analysis had little effect
on their magnitudes and colors --- the most significant change involved
J1009+2656, which became fainter in $g$ by 0.04 mag and redder in $g-r$
by 0.02 mag.

With a median value of $g-r = 0.64$, the total colors of the galaxies in
our sample tend to be fairly red, and most of the objects with red colors
have nuclear spectra that are dominated by light from an old stellar
population (as indicated by the dominant starlight templates used in the
continuum fits).  Only nine galaxies have nuclear spectra whose continua are
associated with either an intermediate-age (e.g., J0811+2328, J0948+0958)
or young (e.g., J1009+2656, J1351+4012, J1405+1146) stellar population.
We note that most of the objects with the lowest [\ion{N}{2}]/H$\alpha$
flux ratios are found amongst this group.  The correspondence between the
line ratios and central star-formation histories of the galaxies might
suggest that nitrogen abundance is a factor.  However, there are several
counter-examples; J1223+5814, J1225+0519, and five other galaxies with low
[\ion{N}{2}]/H$\alpha$ ratios have continua dominated by cool stars, while
J1408$-$0109, which has a very blue nuclear spectrum (fiber $g-r = 0.35$),
has one of the highest [\ion{N}{2}]/H$\alpha$ ratios in the sample.

Despite having a wide range of colors ($g-r =$ 0.27--0.76) and absolute
magnitudes ($M_g = -16.2$ to $-18.7$), the stellar masses of the AGN host
galaxies occupy a fairly narrow range between $\sim 10^9 \, M_{\odot}$
and $10^{10} \, M_{\odot}$.  There are a fair number of objects (9/28) above
$M_{\star} = 8 \times 10^9 \, M_{\odot}$, consistent with the increase in
the numbers of AGNs with increasing stellar mass that we observe in the
full SDSS sample.  At lower masses, the $M_{\star}$ distribution of the AGN
hosts is fairly flat, but the cutoff at $\sim 10^9 \, M_{\odot}$ is abrupt.  
Incompleteness does affect the parent sample below $10^9 \, M_{\odot}$
(see Fig.~1), but there are well over 3000 galaxies in the $10^8 - 10^9 \, 
M_{\odot}$ range, so this cannot be the main reason for the sharp cutoff.  
The argument frequently given for the rarity of emission-line AGNs among
blue late-type galaxies (e.g., Desroches \& Ho 2009) is that such objects
are likely to contain lower mass black holes and have enhanced levels of
star formation.  Thus, we might expect optical signatures of their nuclear
activity to be relatively weak and more easily masked by the emission from
circumnuclear \ion{H}{2} regions.  As Figure~8 shows, the full SDSS sample
does get bluer with decreasing mass, so this scenario may indeed be relevant
for our survey.  On the other hand, some of the AGNs {\it are\/} located
in blue, low-mass galaxies or in higher mass objects with blue central
regions that exhibit evidence of recent star formation.  As we discuss in
\S~5.4, our sample is less affected by luminosity bias than other samples,
so the objects in it may better represent the population of dwarf galaxies
with active nuclei.  If so, the stellar masses and colors of the AGNs shown
in Figure~8 might reflect (in part) the galaxy properties associated with
the presence of a central black hole.  A comparison of the morphologies of
the AGN host galaxies with those of objects from the full SDSS sample with
similar masses would complement the simple color analysis presented here.

\clearpage
\begin{figure}[!htb]
\vskip -0.45truein
\includegraphics[trim=0 100 0 200,clip,width=0.95\textwidth]{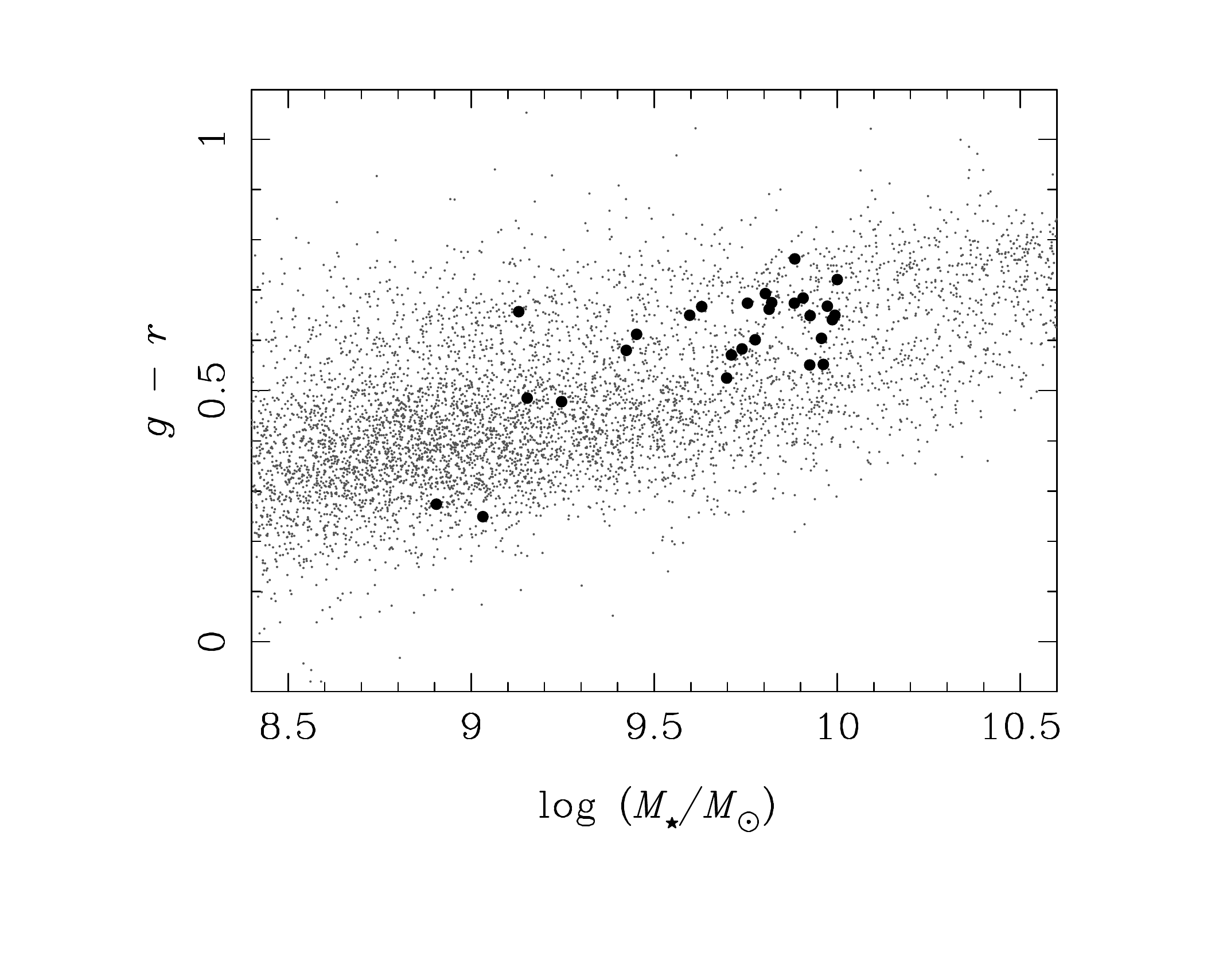}
\end{figure}
\vskip -0.75truein
{\narrower\noindent\small Fig.~8.---Total color vs.\ stellar mass for the full
SDSS sample (grey dots), highlighting the AGNs we have identified (filled
circles).  Galaxies in the parent sample --- as well as the AGN hosts ---
tend to get bluer with decreasing stellar mass.\par}

\subsection{Comparison to Other Searches for Low-Mass AGN Hosts}

One way to assess the results of our survey and the effectiveness of our
approach is to compare our sample of dwarf-galaxy AGNs to those of other
SDSS-based IMBH surveys.  Because of their focus on broad-line AGNs, the
Greene \& Ho (2007b) and Dong et al.\ (2012) surveys are not suitable for
such a comparison.  However, the Barth et al.\ (2008, hereafter BGH08)
study is ideal in several respects.  It, too, targeted low-mass AGN host
galaxies, and as with our sample, most of the objects included in it (27/29)
are narrow-line AGNs.  Most importantly, BGH08 used essentially the same
spectral classification criteria for identifying AGNs, and through detailed
analysis of follow-up high-resolution spectra, they confirmed that all of the
objects in their sample are powered by black-hole accretion.  Statistically,
the main difference between our surveys is in their data selection methods.
To ensure that the comparison is as direct as possible, we have remeasured
all quantities of interest for the BGH08 objects from the SDSS DR7 data in
exactly the same manner as for our own sample, i.e., we determined distances
and host-galaxy photometry the same way, measured emission-line fluxes from
the continuum-subtracted SDSS spectra, and applied all of the various
corrections described above.  The values we obtain for absolute magnitude,
stellar mass, and [\ion{O}{3}] luminosity are thus similar, but not identical,
to those listed in Table~1 of BGH08.

In terms of nuclear luminosity, we find that our sample is substantially
fainter than that of BGH08.  For our sample, $\log (L_{5007}/L_{\odot}$) has
a median of 5.3 and a range of 3.8--6.6, whereas for BGH08 the median and
range are 6.5 and 5.2--7.7, respectively.  As a consequence, the distributions
of minimum black-hole mass (which scales with $L_{5007}$) for the two samples
differ significantly; see Figure~7b.  This could reflect lower actual
black-hole masses in our objects, lower average accretion rates, or some
combination of the two.  A comparison of the host galaxies is not as
straightforward --- stellar masses were computed differently for the BGH08
objects, and they set a different $M_{\star}$ cutoff in their survey.  Still,
the AGN host galaxies in our sample tend to be fainter (median $M_g =$ --17.9,
compared to --18.9 for BGH08) and less massive (median $M_{\star} = 6 \times
10^9\, M_{\odot}$ for our sample vs.\  $1.2 \times 10^{10}\, M_{\odot}$ for
BGH08).  Figure~9, which plots nuclear luminosity and host-galaxy mass,
provides a concise summary of the differences between the samples.  Overall,
the lower values of $L_{5007}$ and $M_{\star}$ of our objects imply that our
sample is less affected by luminosity bias, which is largely due to the
lower distance limit we have adopted for our survey and the different approach
we have used to select AGNs.

Contemporaneous with our work, Reines et al.\ (2013) have presented a sample
of $z \le 0.055$ IMBH candidates in low-mass host galaxies selected from the
SDSS via the NASA-Sloan Atlas.  A comparison between their results and ours
is complicated by differences in the spectral classification criteria, e.g.,
they have included objects with [\ion{O}{3}]/H$\beta < 3$, have segregated
``composite'' objects that fall between the empirical and theoretical maximum
starburst lines on line-ratio plots (see Kewley et al.\ 2006), and have
included a large number of objects that have \ion{H}{2} region-like
narrow-line ratios on one or more of the diagnostic diagrams.  However, most
($\sim$~32/35) of the objects Reines et al.\ classify as ``AGN'' would be
considered Seyferts using our criteria, and likewise, all but two or three
of the 28 objects in our sample would qualify as ``AGN'' according to their
criteria.  Therefore, most of the comparisons that follow pertain to their
``AGN'' subsample.

The nuclear luminosities of the Reines et al.\ AGNs, computed from the
redshifts and [\ion{O}{3}] fluxes listed in their tables, have a range of
$\log (L_{5007}/L_{\odot})$ = 5.2--7.7 (the same as BGH08).  Their median
luminosity exceeds ours by a factor of 5.  In fact, 12 of our objects
have $L_{5007}$ below that of the least luminous Reines et al.\ AGN.  Thus,
as with other SDSS surveys for IMBH candidates, the Reines et al.\ sample
appears to be more affected by luminosity bias than ours.  The reason our
sample extends to lower luminosities is due in part to the fact that we
have not imposed $S/N$ ratio or equivalent width (EW) limits on any of the
diagnostic emission lines, whereas Reines et al.\ required $S/N > 3$ and
EW $> 1$~\AA\ for H$\alpha$, [\ion{O}{3}] $\lambda 5007$, and [\ion{N}{2}]
$\lambda 6583$.  Figure~10 illustrates how these limits would impact our
sample.  As shown in Figure~10a, all of our objects easily exceed the $S/N$
ratio limit for [\ion{O}{3}] $\lambda 5007$ set by Reines et al., but four
of them would fail to meet their [\ion{O}{3}] EW requirement.  As Figure~10b
indicates, these four objects are among our least luminous AGNs.  The
situation is similar for [\ion{N}{2}] $\lambda 6583$ --- three additional
low-luminosity AGNs in our sample that have sufficient $S/N$ ratios would
miss the Reines et al.\ EW cut for this line.  Thus, it would appear that
EW limits, when applied to AGNs that have otherwise well-detected lines,
can contribute to luminosity bias.
\negthinspace\negthinspace\negthinspace
\footnote{Reines et al.\ also required $S/N > 2$ for H$\beta$.  As discussed
in \S~4.3, four AGNs in our sample have H$\beta$ flux upper limits, and given
that all four have low [\ion{O}{3}] luminosities, one could argue that
$S/N$ ratio limits for H$\beta$ are also relevant for luminosity bias.
However, these four objects are among the seven AGNs that would miss
the Reines et al.\ EW cut for one of the other lines.  Thus, the H$\beta$
$S/N$ ratio limit adopted by Reines et al.\ would not by itself cause any
low-luminosity objects to be excluded from our sample.}
Note, however, that five of the 12 objects in Figure~10b with $L_{5007}$ below
the Reines et al.\ minimum would meet all of the emission-line requirements
for inclusion in their survey, so this is not the only reason the luminosity
distributions of our AGN samples differ.

\begin{figure}[!htb]
\vskip 0.05truein
\hskip 0.15truein
\includegraphics[trim=0 200 0 70,clip,width=0.95\textwidth]{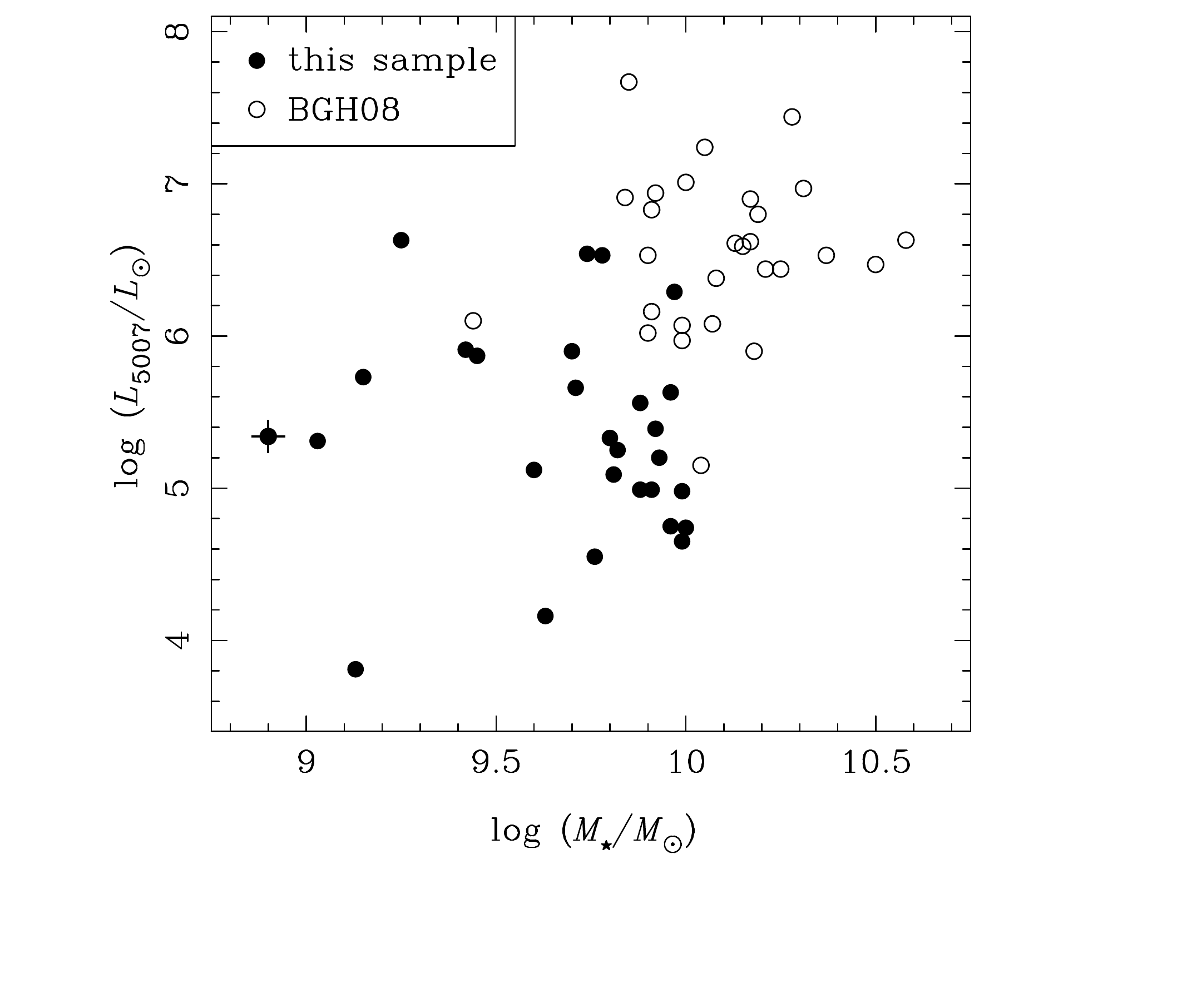}
\end{figure}
\vskip -0.85truein
{\narrower\noindent\small Fig.~9.---Summary of the differences in the nuclear
luminosities (as indicated by $L_{5007}$) and stellar masses of the objects in
our sample and that of BGH08.  The filled circle superposed with a cross marks
the one object common to both samples, J1109+6123.  The generally lower
values of $L_{5007}$ and $M_{\star}$ for our objects suggest that our sample is
less affected by luminosity bias.  The 1~$\sigma$ uncertainties in $L_{5007}$
are of order the size of the symbols shown.\par}
\vskip 0.05truein

Considering stellar masses, Reines et al.\ intended to select objects with
$M_{\star} \le 3 \times 10^9 \, M_{\odot}$.  However, it appears that the SED
fitting technique used to compute their stellar masses yields systematically
lower values of $M_{\star}$ compared to the approach employed here.  Applying
our method to their sample (as we did for BGH08), we find that the stellar
masses of two objects remain the same, but those for the remainder of the
sample all increase by factors of 1.5 to 8.3.  The median increase is a
factor of 2.4; stellar masses range up to $1.1 \times 10^{10} \, M_{\odot}$,
and with one exception --- J1109+6123, which is common to both samples and
whose stellar mass is unchanged --- none of the objects is less massive than
$M_{\star} = 1.0 \times 10^9 \, M_{\odot}$.  Four Reines et al.\ AGNs would
be found in the lowest mass bin ($< 2 \times 10^9 \, M_{\odot}$), compared
to 5/28 for our sample.  This is by no means a statement about which method
for calculating $M_{\star}$ is more accurate --- the point here is that there
is essentially no contrast between our samples in terms their stellar mass
distributions.

The full Reines et al.\ sample of IMBH candidates contains more galaxies
than ours, partly because their survey covers a broader range of
redshifts (80\% of their ``AGNs'' lie beyond our distance limit
\negthinspace\negthinspace
\footnote{In our redshift range, the samples have seven objects in common,
and as discussed in \S~5.2 and \S~5.5, three of these --- J1109+6123,
J1223+5814, and NGC~4395 --- are previously known IMBH candidates.}),
but also because the requirements we have set to establish sample membership
differ considerably.  We estimate that no more than 36\% of their full sample
would meet the AGN selection criteria outlined in \S~5.1.  The types of
objects in the Reines et al.\ sample that we have elected to omit in our
survey are mainly (a) [\ion{N}{2}]/H$\alpha$ ``composite'' galaxies that
have [\ion{O}{3}]/H$\beta$ ratios well below 3, and (b) galaxies whose
narrow-line flux ratios can be fully explained by star-forming activity.

\vskip -0.3truein
\begin{figure}[!htb]
\hskip -0.2truein
\includegraphics[width=0.55\textwidth]{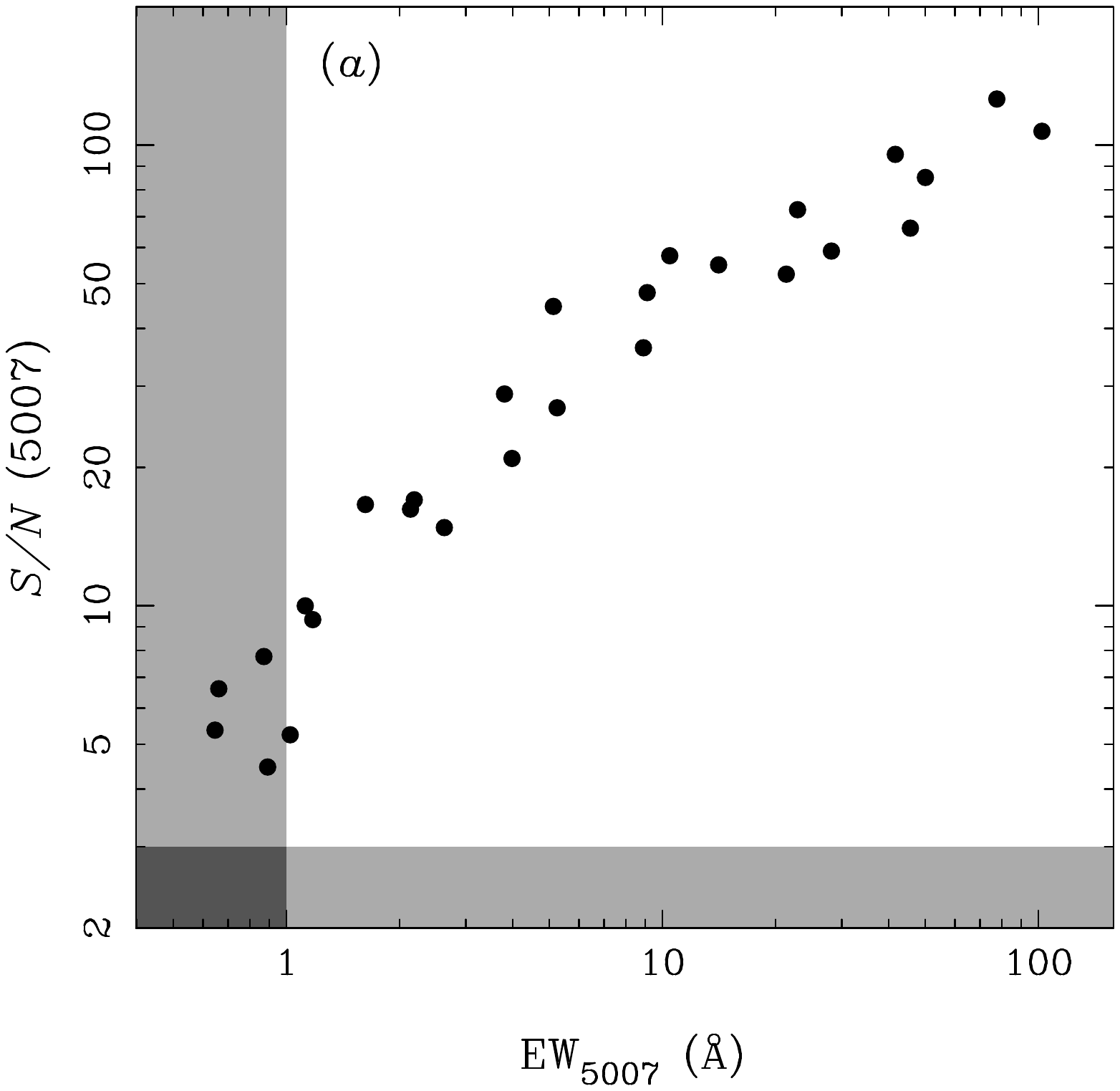}
\vskip -4.645truein
\hskip 3.0truein
\includegraphics[width=0.55\textwidth]{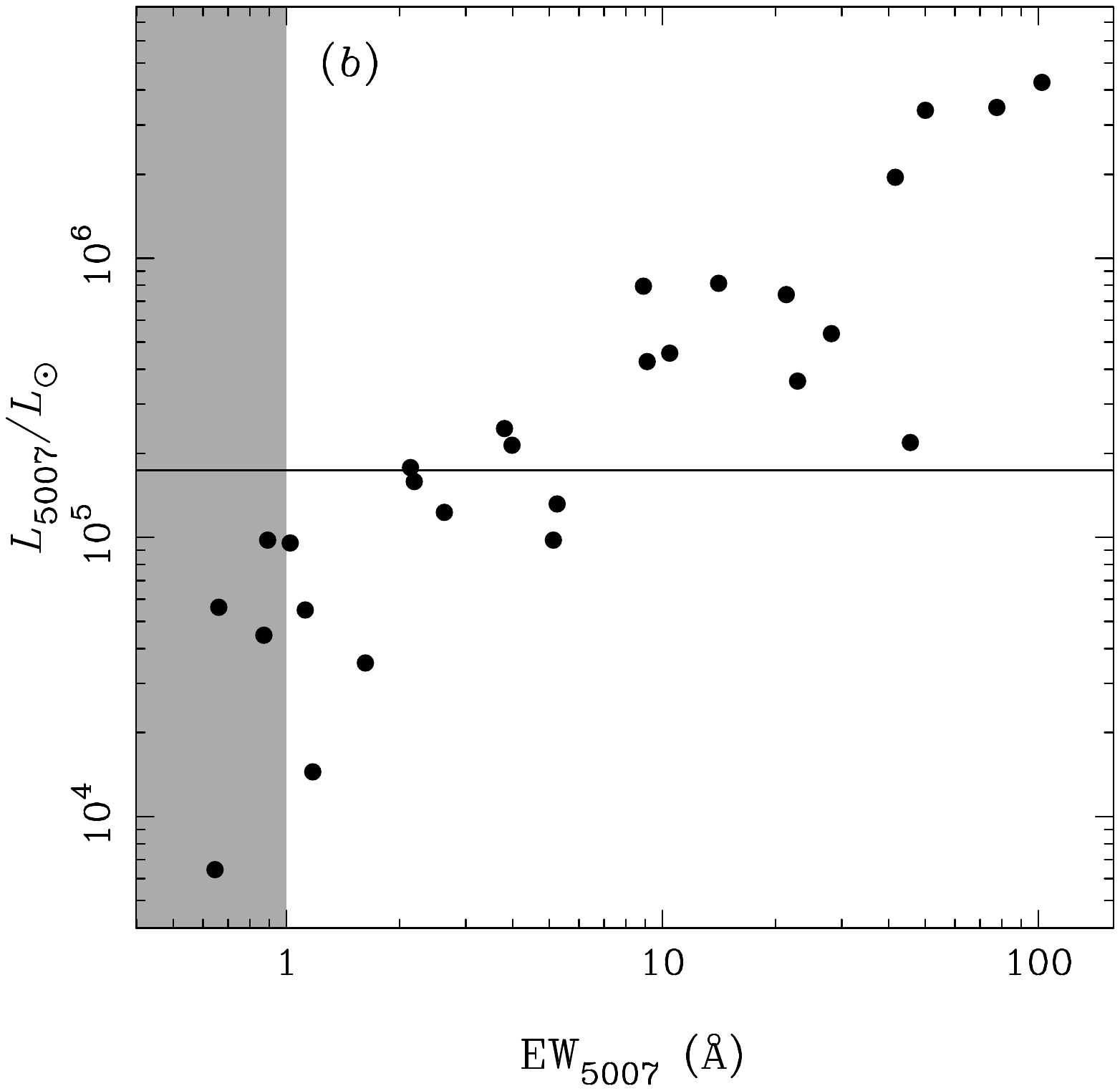}
\end{figure}
\vskip -1.6truein
{\narrower\noindent\small Fig.~10.---($a$) The [\ion{O}{3}]
$\lambda 5007$ $S/N$ ratios (computed by GANDALF in
the full continuum fits) and equivalent widths (EWs) for our sample of AGNs.
The shaded regions indicate the values excluded by Reines et al.\ (2013) in
their search for AGNs in low-mass galaxies.  Four of our objects would
miss their EW cut, but all exceed the $S/N$ ratio limit.  ($b$) Comparison
of the [\ion{O}{3}] luminosities and EWs for our sample.  Again, the
shaded region represents EW values excluded by Reines et al.  The horizontal
line indicates the $L_{5007}$ value of their least luminous AGN.  Applied
to our sample, the EW limit would remove some, but not all, low-luminosity
nuclei.  One object --- NGC~4395 --- has a very high EW$_{5007}$ and is not
included on either plot.\par}

\clearpage
Regarding the low [\ion{O}{3}]/H$\beta$ composites, the majority of such
objects in the Reines et al.\ study mingle with \ion{H}{2} galaxies in
the [\ion{S}{2}]/H$\alpha$ and (when detected) [\ion{O}{1}]/H$\alpha$
diagrams.  Although arguments have been made that this circumstance might
arise from a mixture of star formation and AGN activity (e.g., Kewley et
al.\ 2006), the fact that non-AGN processes can also lead to enhancements
of low-ionization forbidden lines implies that the [\ion{N}{2}]/H$\alpha$
ratio alone is not sufficient to confirm the presence of a massive black
hole in these galaxies.  Many of them --- at least those within our redshift
range that we have examined closely --- have low emission-line luminosities
that could be powered by stellar processes (e.g., photoionization by evolved
stars; see Binette et al.\ 1994).

We also have excluded ambiguous objects that, when all available spectral
evidence is considered, are more likely to be star-forming galaxies than
AGNs.  Objects in this category often have very weak low-ionization forbidden
lines, but their high [\ion{O}{3}]/H$\beta$ ratios place them close to (or
just over) the maximal starburst lines on diagnostic diagrams.  In some cases,
the ratios may be consistent with those of low-metallicity AGNs.  But in our
sample, all such objects clearly have stellar continua in both the optical,
where O and B stars dominate, and the far UV, where, based on their low
starburst-like \ion{He}{2} $\lambda 4686$ fluxes, a very steep continuum
slope of $\alpha_{\rm UV} = 4$--5 is indicated.  As the helium abundance has
only a weak dependence on metallicity (Groves et al.\ 2006), we expect that
the \ion{He}{2}/H$\beta$ ratios of true low-metallicity AGNs should exceed
those of starbursts.  A few objects with these spectral characteristics
are tentatively identified as pure AGNs in the Reines et al.\ (2013)
sample (e.g., J0823+0313 and J0840+4707).

The other class of star-forming galaxies that we have excluded are those
with starburst-like narrow-line flux ratios and broad H$\alpha$ emission.
In principle, some of these could be extreme composite objects where the
broad Balmer lines are the only AGN features not masked by the starburst.
However, this is by no means the only possible interpretation, as some
nearby extragalactic \ion{H}{2} regions are known to display broad H$\alpha$
emission.  One example is NGC~5471, which is located $\sim 25$ kpc from
the center of M~101 and is not suspected of harboring an IMBH.  Nevertheless,
the luminosity and velocity width of the H$\alpha$ line in this object
($L_{{\rm H}\alpha} = 2 \times 10^{38}$ ergs~s$^{-1}$ and FWHM(H$\alpha$) =
1500 km~s$^{-1}$; Isotov et al.\ 2007) imply a virial black-hole mass (via
Xiao et al.\ 2011) of $\log (M_{\rm BH}/M{\odot}) = 5.1$, which would
represent 10--50\% of the mass of the entire stellar cluster in NGC~5471
(Garc\'ia-Benito et al.\ 2011).  Along similar lines, some of the star-forming
galaxies with broad H$\alpha$ emission in the Reines et al.\ (2013) sample
(e.g., objects B, D, and H in their Tables 3 and 6) have estimated black-hole
masses that are $\sim 1$\% of the total stellar mass of the host galaxy --- 
potentially very important in terms of galaxy/black hole co-evolution, or
an indication that the broad H$\alpha$ emission is unrelated to the presence
of a black hole.  In the absence of other AGN indicators, vigorous star
formation seems to offer the more straightforward explanation for such objects.

\subsection{Notes on Individual Objects}

{\it J0948+0958}: This galaxy has two superposed foreground stars and it
lies $\sim 1'$ from a very bright star.  We also noticed that the model
magnitudes are significantly brighter ($\sim 0.4$ mag in $g$) than the
Petrosian magnitudes.  We thus performed our own aperture photometry of the
galaxy in the SDSS images with the {\it phot\/} task in IRAF, using a local
background (measured at the same radius from the bright star as the galaxy)
and removing the contribution of the foreground stars.  We obtained
magnitudes in $g$, $r$, and $i$ that are within a few hundreths of a
magnitude of the model magnitudes for the galaxy, so we have adopted the
latter for our calculations.

{\it J0949+3213}: This object, also known as NGC~3011 and (more noteworthy)
Mrk~409, is included in the Markarian survey of ultraviolet-excess galaxies.
Interestingly, it has been classified repeatedly as a starburst galaxy in
the literature (e.g., Balzano 1983; Salzer et al.\ 1995).  Indeed, the
object has a prominent blue star-forming ring located just outside the
nucleus (Gil de Paz et al.\ 2003; see our Fig.~4).  However, the SDSS
spectrum leaves no question about the nature of the nuclear activity of this
galaxy. In addition to narrow emission-line flux ratios that place it well
within the Seyfert regions on the diagnostic diagrams in Figure~5 (particularly
on the [\ion{O}{1}]/H$\alpha$ plot), the object has strong \ion{He}{2}
$\lambda 4686$ emission.

{\it J1109+6123}: As mentioned above, this is the one galaxy that is common to
our sample and that of Barth et al.\ (2008).  Note that because our methods
for estimating distances, stellar masses, and emission-line luminosities are
not identical, the parameters listed in Tables~1 and 2 differ somewhat with
those published by Barth et al.  Our value of $M_{\star}$ agrees closely with
the revised stellar mass computed by Thornton et al.\ (2009).

{\it J1207+4307}: Also known as NGC~4117, this object was identified early
on as a Seyfert galaxy (Huchra et al.\ 1982) despite having a rather low
(compared to classical Seyferts) [\ion{N}{2}]/H$\alpha$ ratio.  In our
sample, it has the weakest detected \ion{He}{2} emission, but its [\ion{O}{1}]
strength (relative to H$\alpha$) is second only to that of NGC~4395.  In fact,
on the [\ion{O}{1}]/H$\alpha$ plot in Figure~5, NGC~4117 falls on the boundary
between Seyferts and LINERs proposed by Kewley et al.\ (2006).  It is, however,
a bright X-ray source with an absorption column density ($N_{\rm H} = 4 \times
10^{23}$~cm$^{-2}$; Cardamone et al.\ 2007) similar to that commonly observed
in the X-ray spectra of luminous type~2 Seyfert nuclei.  Thus, there is no
question that this object is powered by black-hole accretion.

{\it J1225+0519}: This galaxy formally falls within the coordinate boundaries
and velocity range often used to define membership in the Virgo cluster
(Mould et al.\ 2000).  Hence, it is also known as VCC~764.  However, if the
galaxy were in the cluster and at a distance of $\sim 14$ Mpc, it would have
by far the faintest absolute magnitude and lowest stellar mass of any object
our AGN sample.  Given that the SDSS redshift is close to the upper limit
assumed for Virgo membership, we considered the possibility that the galaxy
is instead behind Virgo and computed its infall-corrected distance
accordingly.  This distance, 33.4 Mpc, implies a nuclear luminosity and
host-galaxy properties that are much more in line with the rest of our
sample, so we have adopted it here.

{\it J1225+3332}: This galaxy (NGC 4395), being the nearest and (in an
angular sense) most extended object in our sample, requires special
consideration in terms of its distance, photometry, and spectral analysis.
We adopt the Cepheid distance of 4.3 Mpc for this galaxy (Thim et al.\
2004), which is far more accurate than its redshift-based distance.  Also,
because the object is both extended and knotty, all Petrosian and model
magnitudes listed in the DR7 and DR8 are much too faint.  We have instead
used the $B$ and $V$ magnitudes from the RC3 (de~Vaucouleurs et al.\ 1991)
and 2MASS $K$ magnitude (Jarrett et al.\ 2003) to estimate its host-galaxy
properties.  We calculated the galaxy's stellar mass using the $BVK$
magnitudes and the appropriate coefficients from Bell et al.\ (2003).
The absolute $g$-band magnitude listed in Table~1 was computed from these
data using the transformations of R.~Lupton (2005).
\negthinspace\negthinspace\negthinspace
\footnote{See http://www.sdss.org/dr7/algorithms/sdssUBVRITransform.html}
The same transformations
imply a $g-r$ color of 0.249, which we used to determine the location of
this object in Figure~8.

Also, because the galaxy is so nearby and has a faint central surface
brightness, the SDSS spectrum appears to contain light only from the
nucleus.  It is the only object in our sample for which a pure power
law was used to model the continuum.

{\it J1351+4012}: As discussed above, this object is a good candidate for
a Seyfert/starburst composite galaxy based on its emission-line flux ratios.
Similar to J0949+3213, it is a UV-excess Markarian galaxy (Mrk 462) with
a compact, blue central region that suggests active star formation.  However,
as noted in \S~5.1, other spectral evidence points strongly to the presence
of an accreting black hole.  Its only published spectral classification 
is that of ``emission spectrum galaxy'' (Petrosian et al.\ 2007).

{\it J1448+1227}: This galaxy, also known as NGC~5762, has been studied
because of its 21~cm neutral hydrogen emission (Lewis et al.\ 1985;
Rosenberg et al.\ 2000) and active star-formation (e.g., Kewley et al.\ 2002;
note its blue, \ion{H}{2} region-studded disk in Fig.~4).  Interestingly,
little attention has been given to its nuclear properties.  The continuum
in the SDSS spectrum is dominated by light from an older stellar population, 
and the emission-line flux ratios are clearly those of a type~2 Seyfert.

{\it J1605+0850}: Although the DR7 Petrosian and model magnitudes for
this galaxy are similar, the radius of the region used for the aperture
photometry is much smaller than its total optical extent ($\sim 4''$ vs.\
$\sim 12''$), meaning all of the DR7 magnitudes are too faint.  As with
J0948+0958, this galaxy lies within $1'$ of a bright foreground star,
which may have affected the sky subtraction.  Regardless, the problem
seems to have been corrected in the DR8 --- our manually measured magnitudes
(following the method described above for J0948+0958) agree well with the
DR8 Petrosian magnitudes, so we have adopted the latter for our analysis.

\section{Summary and Conclusions}
Nearby dwarf galaxies with intermediate-mass black holes afford a glimpse
into the earlier stages of black hole/galaxy co-evolution.  Owing to the
fact that both components have undergone relatively little growth over
cosmic time, dwarf galaxy AGNs also provide an opportunity to discriminate
between models for the formation of massive black-hole seeds --- a population
which, at present, does not lend itself to direct observational study.
Using a distance-limited portion of the SDSS DR7, we have assembled a sample
of 28 AGNs within $\sim 80$ Mpc that reside in galaxies with stellar
masses less than $10^{10}\, M_{\odot}$ (calculated via the Bell et al.\ 2003
method).  As we have excluded emission-line galaxies that could be powered
by stellar processes, these objects are the least massive galaxies in the
very local universe certain to contain central black holes.  We estimate
that $\sim 75$--80\% of all nearby galaxies with similar absolute magnitudes
($M_g \approx$ --17 to --18) have SDSS spectra (E.~Moran et al.\ 2015, in
preparation), implying that the completeness of our optically selected
sample is high.

We began the process of constructing this sample by treating every galaxy
with a nuclear SDSS spectrum that falls within our adopted distance limit
as a potential AGN host.  We then carefully modeled the stellar continua
of the objects, combining automatic emission-line measurements with visual
assessments of the continuum-subtracted spectra, to classify the type of
activity present.  By focusing on the nearest galaxies and not imposing
explicit $S/N$ ratio or equivalent width limits on the diagnostic emission
lines, this approach has yielded a sample of AGNs with very low luminosities
and host-galaxy masses.  As a result, our survey is less affected by
luminosity bias than previous SDSS searches for IMBH candidates.  The low
fraction of objects with previous X-ray or radio detections suggests that
optical spectroscopy is the most economical way to identify low-luminosity
active nuclei in dwarf galaxies.

Based on the SDSS spectra, the vast majority of AGNs in our sample appear
to be narrow-line (type 2) objects.  Although the non-uniform quality of
the data prevents us from drawing firm conclusions about the true ratio
of narrow- and broad-line AGNs at low luminosities, there is nothing about
our selection methods that would have caused us to miss obvious type~1
objects.  We are currently obtaining high $S/N$ ratio follow-up spectra,
in part so that we may place tighter constraints on the broad-line AGN
fraction for this sample.  In addition, we have obtained X-ray data for
a subset of the objects to search for evidence of heavy absorption in
the type~2 objects similar to that observed in classical Seyfert~2 nuclei.
Obscured or not, the low nuclear luminosities displayed by our sample
imply low minimum black hole masses in the $\sim 10^3 - 10^4$ $M_{\odot}$
range, which is approximately where the initial mass functions predicted for
different black-hole seed formation scenarios overlap the most (e.g.,
Volonteri 2010).

Although many of the host galaxies of the AGNs we have found appear to be
disk-dominated systems, most of them are not blue, dwarf spirals similar
to the original IMBH host, NGC~4395.  One possibility is that, at the lower
nuclear luminosities exhibited in our sample, we include more typical
members of the IMBH population, which in turn are found in the more typical
host galaxies of such objects.  Alternatively, the low incidence of vigorous
star formation amongst the AGN hosts might be a reflection of the fact that,
at low luminosity, the emission-line signatures of black-hole accretion are
more easily masked by emission from circumnuclear star-forming regions.
If so, high spatial resolution spectroscopy of blue, low-mass galaxies
(especially those that display ambiguous features in their SDSS specrta)
and searches for non-optical AGN emission in these objects could provide
a more complete census of IMBHs and their host galaxies.

There is a general impression that AGNs in low-mass galaxies are rare (e.g.,
see Kauffmann et al.\ 2003).  If this is the case, then new, nearby examples
such as those presented here will be valuable additions for investigations
into the relationship between black holes and galaxies.  However, we
note that an appearance of rareness is likely to occur in AGN samples that
are biased in favor of high-luminosity objects; AGN fractions derived from
surveys that emphasize low-luminosity objects should be more accurate.  In
this study, if we examine just the $M_{\star} = (4-10) \times 10^9$ $M_{\odot}$
range --- which is free of significiant incompleteness (Fig.~1b) and
includes 75\% of the AGN host galaxies in our survey (Fig.~8) --- we obtain
an AGN fraction of 2.7\%.  This fraction increases to 5.0\% if we only
consider objects in the color range exhibited by the AGNs (i.e., $g - r =
0.5 - 0.8$).  The latter figure is probably optimistic, but not entirely
so if circumnuclear star formation has affected our ability to detect AGNs
in bluer galaxies.  Although much more work lies ahead --- developing truly
complete samples and a gaining a better handle on the limitiations of optical
spectroscopy --- AGN fractions in this range are becoming relevant in terms
of the predicted black-hole occupation fractions expected for galaxies of
this mass.

\acknowledgments
Generous support for this work was provided by the National Science Foundation
through grant AST-0909063.  We are very grateful to the referee, Jenny Greene,
for her thorough and insightful review of this article, which helped us to
make a number of valuable improvements.  We would also like to thank Michele
Cappellari for providing access to the pPXF code, and Marc Sarzi for sharing
the GANDALF software and for his time during numerous conversations regarding
its inner workings.  This research has made use of a number of other publicly
available resources, for which we are grateful: synthethic stellar population
models from the MILES team; the NASA/IPAC Extragalactic Database (NED), which
is operated by the Jet Propulsion Laboratory, California Institute of
Technology; the $K$-correction calculator of I.~Chilingarian and colleagues;
and of course, the photometric and spectroscopic databases from the SDSS
collaboration. E.C.M.\ would like to express special thanks to Meg Urry and
the staff of the Yale Center for Astronomy and Astrophysics for their
hospitality during a sabbatical visit, when much of the work presented here
was completed.  M.E.\ would like to thank the Center for Relativistic
Astrophysics at Georgia Tech and the Department of Astronomy at the University
of Washington for their warm hospitality during the last stages of his work
on this paper.

\end{document}